# Basics of averaging of the Maxwell equations for bulk materials

A. Chipouline, C. Simovski, S. Tretyakov

## Abstract

Volume or statistical averaging of the microscopic Maxwell equations (MEs), i.e. transition from microscopic MEs to their macroscopic counterparts, is one of the main steps in electrodynamics of materials. In spite of the fundamental importance of the averaging procedure, it is quite rarely properly discussed in university courses and respective books; up to now there is no established consensus about how the averaging procedure has to be performed. In this paper we show that there are some basic principles for the averaging procedure (irrespective to what type of material is studied) which have to be satisfied. Any homogenization model has to be consistent with the basic principles. In case of absence of this correlation of a particular model with the basic principles the model could not be accepted as a credible one. Another goal of this paper is to establish the averaging procedure for bulk MM, which is rather close to the case of compound materials but should include magnetic response of the inclusions and their clusters. In the vast majority of cases the consideration of bulk materials means that we consider propagation of an electromagnetic wave far from the interfaces, where the eigenwave in the medium has been already formed and stabilized. In other words, in this paper we consider the possible eigenmodes, which could exist in the equivalent homogenized media, and the necessary math apparatus for an adequate description of these waves.
A discussion about boundary conditions and layered MM is a subject of separate publication and will be done elsewhere.

## Content

## 1 Homogenization of Maxwell equations – macroscopic and microscopic approaches

### 1.1 Microscopic Maxwell equations and averaging procedure

We consider as a starting point a system of microscopic MEs in the following form:

$$\begin{cases} \mathrm{rot}\,\vec{e} = \dfrac{i\omega}{c}\vec{h} \\ \mathrm{div}\,\vec{h} = 0 \\ \mathrm{div}\,\vec{e} = 4\pi\rho \\ \mathrm{rot}\,\vec{h} = -\dfrac{i\omega}{c}\vec{e} + \dfrac{4\pi}{c}\vec{\mathrm{j}} \end{cases} \qquad \begin{cases} \rho = \sum\limits_i q_i \delta\left(\vec{r} - \vec{r}_i\right) \\ \vec{\mathrm{j}} = \sum\limits_i \vec{v}_i q_i \delta\left(\vec{r} - \vec{r}_i\right) \\ \dfrac{d\vec{p}_i}{dt} = q_i \vec{e} + \dfrac{q_i}{c}\left[\vec{v}_i * \vec{h}\right] \end{cases} \qquad (1)$$

Here $\vec{e}$ and $\vec{h}$ are the microscopic electric and magnetic fields, respectively, $\rho$ is the charge density, $\vec{q_i}$, $\vec{p_i}$, $\vec{r_i}$ and $\vec{v_i}$ are the charges, pulses, coordinates and velocities of charges, $\vec{j}$ is the microscopic current density, $\omega$ and $c$ are the frequency and the velocity of light in vacuum. It is assumed that system (2.1) is strictly valid without any approximations. Actually, system (1) can be elaborated in the framework of the minimum action approach [Landau&Lifshits]; nevertheless, one should remember that the minimum action principle does not give an unambiguous form of the MEs (1), but instead gives a set of different forms which satisfy the requirement of relativistic invariance. The “right” form can be chosen based on the evident requirement of correspondence of the results of the final system of equations to the observed physical effects. One should also mention that in the framework of the minimum action approach the final equations are written for “potentials + particles”, not for “fields + particles”; the respective equations are:

$$\begin{cases} \dfrac{d\overrightarrow{\mathrm{p}_i}}{dt} = -\dfrac{q}{c}\dfrac{\partial \overrightarrow{\mathrm{A}}}{\partial t} - q_i \nabla\varphi + \dfrac{q_i}{c}\left[\vec{v}_i * \mathrm{rot}\,\overrightarrow{\mathrm{A}}\right] \\ \dfrac{\partial \mathrm{F}^{ik}}{\partial x^k} = -\dfrac{4\pi}{c}\mathrm{j}^i, \quad \mathrm{F}_{ik} = \dfrac{\partial A_k}{\partial x_i} - \dfrac{\partial A_i}{\partial x_k} \end{cases} \tag{2}$$

Here $\vec{A}$ and $\varphi$ are the components of the 4-vector potential, and the relations between the microscopic fields and the potentials are given by:

$$\begin{cases} \vec{e} = -\dfrac{1}{c}\dfrac{\partial \overrightarrow{\mathrm{A}}}{\partial t} - \nabla\varphi \\ \vec{h} = \mathrm{rot}\,\overrightarrow{\mathrm{A}} \end{cases} \tag{3}$$

Form (2) will not be used in the following discussions and is presented here just for methodological reasons.

It has to be mentioned that the basic formulation of electrodynamics is still under discussion [Dubovik 93]. Here it is assumed that system (1) fully describes the electromagnetic phenomena and further discussion about validity of (1) is left out.

We consider propagation of an electromagnetic plane wave interacting with the medium in case when the classical dynamics is supposed to be valid and the bulk material fills the whole space; the system (1) in this case can be formally averaged over a physically small volume (or through statistic averaging), which results in:

$$\begin{cases} \operatorname{rot}\left\langle \vec{e} \right\rangle = \dfrac{i\omega}{c}\left\langle \vec{h} \right\rangle \\ \operatorname{div}\left\langle \vec{h} \right\rangle = 0 \\ \operatorname{div}\left\langle \vec{e} \right\rangle = 4\pi\left\langle \rho \right\rangle \\ \operatorname{rot}\left\langle \vec{h} \right\rangle = -\dfrac{i\omega}{c}\left\langle \vec{e} \right\rangle + \dfrac{4\pi}{c}\left\langle \vec{\mathrm{j}} \right\rangle \\ \rho = \sum\limits_i q_i \delta\left(\vec{r} - \vec{r}_i\right) \\ \left\langle \vec{\mathrm{j}} \right\rangle = \sum\limits_i q_i \left\langle \vec{v}_i \delta\left(\vec{r} - \vec{r}_i\right)\right\rangle \\ \dfrac{d\vec{v}_i}{dt} = \dfrac{q_i}{m_i}\vec{e} + \dfrac{q_i}{m_i c}\left[\vec{v}_i * \vec{h}\right] \end{cases} \xrightarrow[\left\langle \vec{h} \right\rangle = \vec{B}]{\left\langle \vec{e} \right\rangle = \vec{E}} \begin{cases} \operatorname{rot}\vec{E} = \dfrac{i\omega}{c}\vec{B} \\ \operatorname{div}\vec{B} = 0 \\ \operatorname{div}\vec{E} = 4\pi\left\langle \rho \right\rangle \\ \operatorname{rot}\vec{B} = -\dfrac{i\omega}{c}\vec{E} + \dfrac{4\pi}{c}\left\langle \vec{\mathrm{j}} \right\rangle \\ \left\langle \rho \right\rangle = \sum\limits_i q_i \delta\left(\vec{r} - \vec{r}_i\right) = \left\langle \rho \right\rangle\left(\vec{E}, \vec{B}\right) \\ \left\langle \vec{\mathrm{j}} \right\rangle = \sum\limits_i q_i \left\langle \vec{v}_i \delta\left(\vec{r} - \vec{r}_i\right)\right\rangle = \left\langle \vec{\mathrm{j}} \right\rangle\left(\vec{E}, \vec{B}\right) \end{cases} \quad (4)$$

The averaging is usually performed in case of a large number of atoms/molecules in the volume of averaging; from the other side the volume is supposed to be small in comparison with the wavelength of the electromagnetic wave, propagating in the medium.

*The main problem here is to find the averaged current and charge distribution as functions of the averaged electric and magnetic fields:*

$$\begin{aligned} \left\langle \vec{\mathrm{j}} \right\rangle &= \left\langle \vec{\mathrm{j}} \right\rangle\left(\vec{E}, \vec{B}\right) \\ \left\langle \rho \right\rangle &= \left\langle \rho \right\rangle\left(\vec{E}, \vec{B}\right) \end{aligned} \quad (5)$$

Averaged (or macroscopic) fields result from the averaging of the actual, microscopic fields which are produced by some external sources and charged particles. The averaging procedure itself is rarely considered in phenomenological models of macroscopic Maxwell equations. In order to set the applicability limitations for the averaging procedure, it is necessary to determine the procedure itself. For example, in the frame of the Lorenz-Lorentz concept, the relation between the wavelength and averaging volume is chosen so that the particle size and the distances between them are 2-3 orders of magnitude smaller, than the wavelength (in Fig. 1 this means $L_{inter}, L_{intra} \leq 10^{-2\div-3}\lambda$ ). This requirement fails in case of composites even for UH frequencies, but nevertheless the Lorentz-Lorentz approach still gives very good results. It turns out that the requirement $L_{inter}, L_{intra} \leq 10^{-2 \div -3}\lambda$ can be significantly relaxed, and the main question in this case becomes: up to which relations between particle size and the wavelength the averaging procedure will still make sense. First, the averaging volume (typical size of this volume) has to be in any case less than the wavelength, otherwise the field (which is supposed to be harmonic in space) disappears. Roughly speaking, the typical size of the averaging volume has to be in any case 5-10 times smaller than the wavelength. Taking into account that the particle size and the distance between particles have to be only order of magnitude less, than the wavelength, it results in no more than one particle per averaged volume. Hence, the averaging concept for this case is required to be qualitatively different in compare with the classical (for example, Lorentz-Lorentz) one.

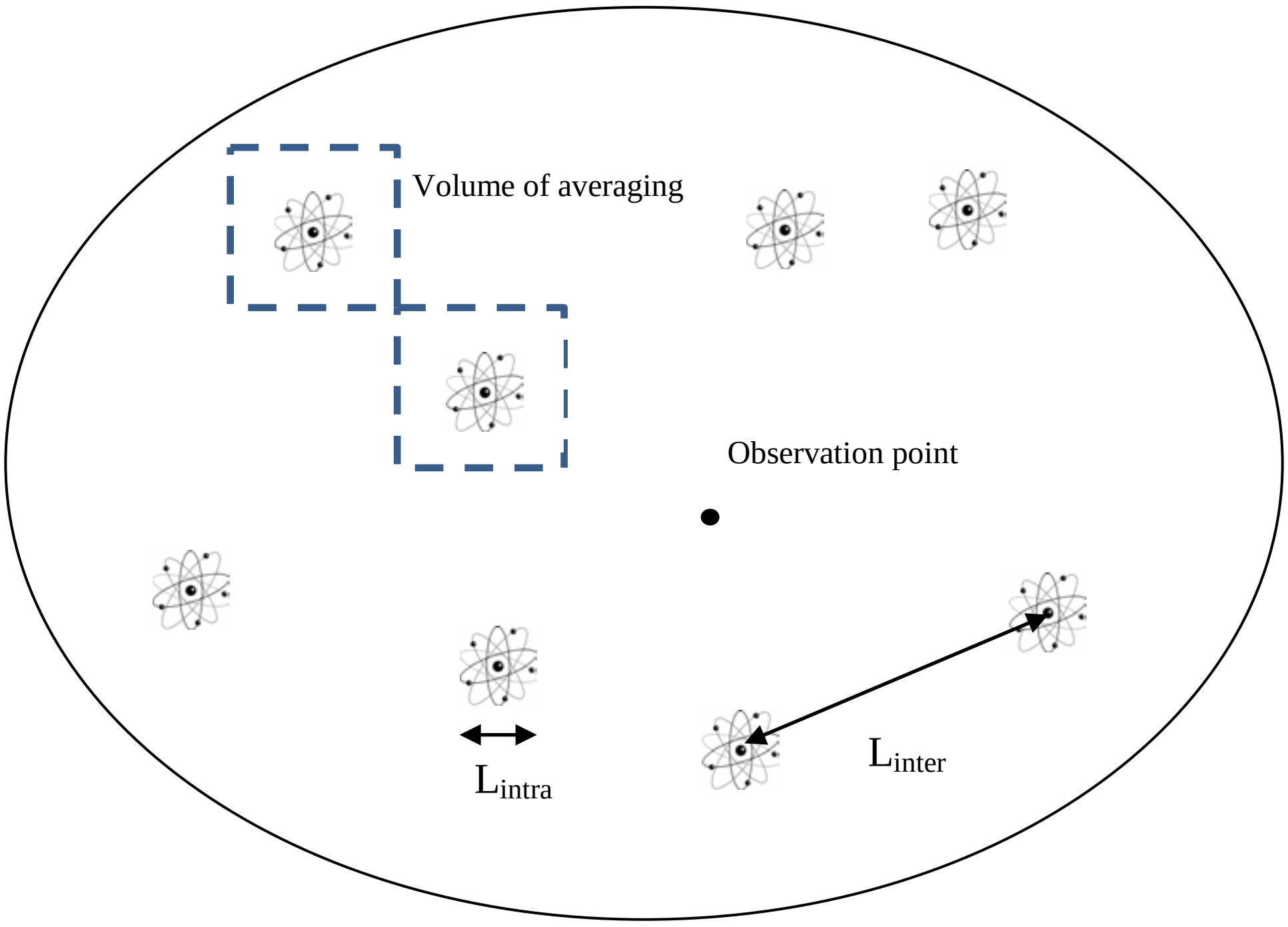

**Fig. 1: Schematic representation of the media which is subject of the averaging with typical lengths, involved in consideration**

The key notion for the introduced averaging procedure is the volume of averaging – the unit cell of the procedure (see Fig. 1).  All volume is subdivided into cells (the "volume of averaging" in Fig 1), and each of them contains at least one particle. Moreover, it is supposed that the composite is homogeneous, i.e. there are no empty cells and there are no cells containing more than one particle. It means that the considered media is regular, but the particle in the cell is not necessary positioned at the center of the cell; each cell contributes to the fields at the point of consideration, from which all averaged characteristics are supposed to be depended on. The average volume is taken equal to the volume of the elementary cell (has been first proposed in [Schwinger 98]).

There are two main approaches to the averaging, namely statistical one [Mazur 53], [Rusakoff 70], [Maradudin 73], where averaging is taking over the ensemble of realization; and spatial, where the main questions are volume of the averaging and the averaging function, which set the minimal macroscopic scale at which change of the macroscopic (averaged) functions is still significant [Jackson 99].

A significant advantage of the statistical approach is the absence of characteristic scales, like the volume of the unit cell, because the averaging is performed over all possible realisations rather than over physical volume. A drawback of the statistical approach is in relatively complex math required for the elaboration of the model. The concept of the statistical averaging, being well developed for the electrostatics and magnetostatics, appears not to be completed for the microwave frequencies and optical spectra [Maslovsky 04].

One more approach, called scaling algorithm has to be mentioned [Vinogradov 99]. This is a generalized method of spatial averaging, where at the first step the averaging is performed over small-scale cells, then over bigger cells, which include many small cells. The procedure is repeated before the averaged functions converge to their asymptotical values. The method is based on specially introduced multipoles (different from the usually defined ones), which are assigned to the cells at each step of the averaging. The method (similarly to [Rusakoff 70]) is a universal one for diluted and dense composites, because it does not require introduction of the local field, but is relatively complicated and not constructive, e.g. the method itself does not suggest constructive averaging procedure.

More comprehensive general review of different approaches to the problem of averaging can be found in [Simovski 03].

### 1.2 System under consideration

In spite of the fact, that in this work general properties of ME (irrespective to a particular medium) are of interest, it seems to be methodologically appropriate to determine from the beginning the type of metamaterials (MM) which will be considered and keep this type in mind throughout the text.

In this work a medium consisting of artificial metaatoms embedded in a dielectric matrix will be considered – see Fig. 2, where only one layer of the considered material is presented.

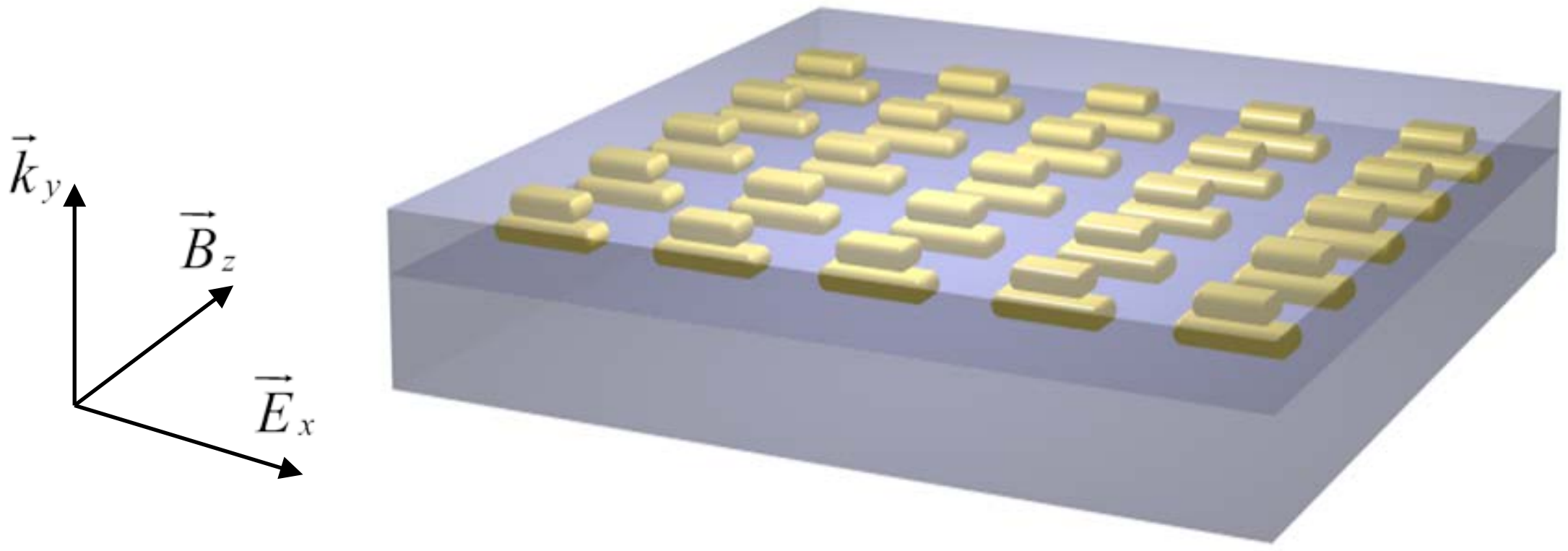

**Fig. 2: Artificial metaatoms (plasmonic nano resonators) embedded in a dielectric matrix form a MM (only one layer is presented). Polarization of the electric and magnetic fields, and direction of the wave vector are shown.**

The metaatoms are assumed to be complex plasmonic structures, possessing so called symmetric and antisymmetric eigen modes - see Fig. 3, where one possible structure (coupled nanowires, in general of different sizes) is shown. The structure consists of two nanowires with typical lengths (for optical domain) of tens to hundreds nanometers, placed one under another with the distance of several tens of nanometers, ensuring strong near field interaction between both nanowires [Svirko 01]. Considering both nanoresonators as two coupled harmonic oscillators, it becomes clear that the structure possess two fundamental eigen modes, namely symmetric (Fig. 3-a) which produces effective dielectric response, and antisymmetric (Fig. 3-b) which is responsible for the magnetic response of the media.

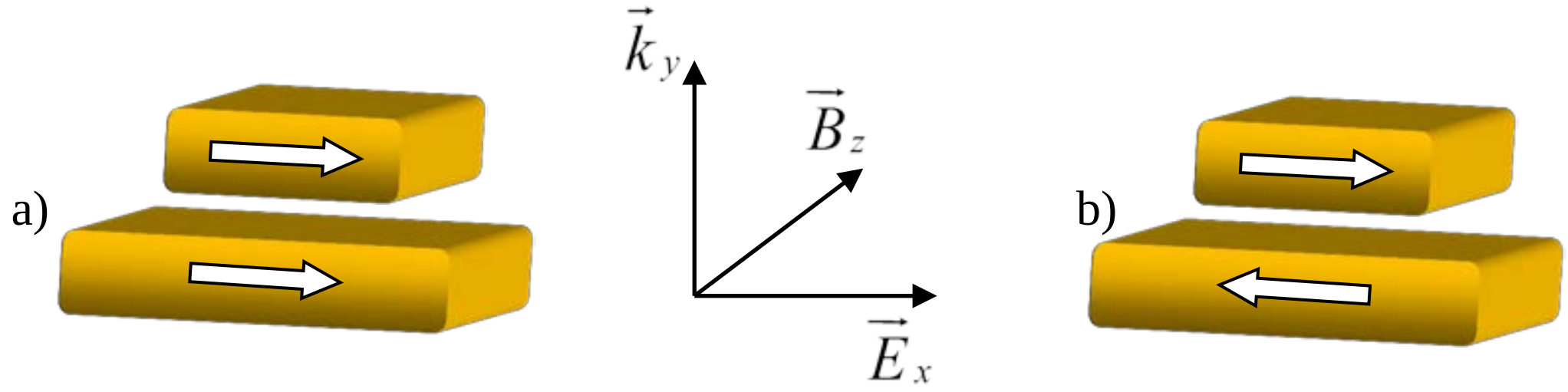

**Fig. 3: One of the possible shape of metaatoms, possessing a) symmetric and b) anti symmetric modes. Electric field $\vec{E}_y$ of the incoming wave, propagating along $y$ axis excites eigen modes of the plasmonic metaatom.**

Here the incoming electromagnetic wave interacts with the electrons of the plasmonic structure and effectively excites symmetric and asymmetric oscillations, provided that the frequency of the incoming wave is close to the respective eigen frequency of the eigen modes. In case of symmetric mode electrons in both nanoresonators oscillate in phase, while in case of antisymmetric mode – out of phase (Fig. 3-a and Fig. 3-b, respectively).

It is also intuitively clear, that in case of symmetric oscillation (Fig. 3-a) the metatom formed by two wires of equal length does not produce any magnetic effects, but rather exhibits extra dipole moments leading finally to the change of the permittivity of the MM. The main interest to the MM is stipulated by the possibility to excite the antisymmetric modes (Fig. 3. b)). In this case the structure presents to the first approximation a circle current, which provides (as it is well known from the school course of physics) a magnetic response. The fantastic peculiarity of the MMs is in the fact, that the MMs provide magnetic response at optical frequencies, where no natural media has similar properties. This fundamentally distinguishes MMs from any natural materials, and makes such MMs (among others) extremely interesting objects for both fundamental research and various applications [Zheludev Roadmap].

In most parts of this work, a MM consisting of coupled nanowires will be considered. In spite of the wide range of different shapes of the nanoresonators considered in the publications (including extremely exotic ones [Papasimakis 09]), the coupled nanowire was the first structure whom a negative refractive index has been demonstrated with [Grigorenko 05], [Shalaev 05]. Moreover, properties of this structure allow clear physical interpretation and relatively simple analytical treatment, which makes coupled nanowires a very good object for discussion of different physical models. One of the simple and at the same time rather good examples is in consideration of necessary conditions for the existence of antisymmetric modes. Remembering the mentioned above interpretation of the dynamics of the coupled nanowires in terms of coupled harmonic oscillators, it is clear that the antisymmetric mode can be excited due to the following reasons:

- First, asymmetric excitation (for example, retardation at the wave propagation between lower and upper nano wires), and,
- Second, due to asymmetric shape of the structure (not equal sizes of the upper and lower nanowires).

Both cases lead to rather different optical properties of the metamateris consisting of such metaatoms.

It has to be also emphasized that the presented work is concentrated mainly on homogenization of bulk material – multilayer (in Fig. 2 only one layer is presented) MM.

Comparing Fig. 1 with Fig. 2 and identifying the metaatoms as artificial atoms one can immediately conclude, that the basic requirements for the homogenization procedure are satisfied rather poorly. For example, typical sizes of the metaatoms, having the resonance wavelength of about 1 μm, are of the order of several hundreds of nanometers; the distances between the metaatoms are of the same order. It is clear, that classical approach for homogenization which results in known expressions like ones for small concentrations and small permittivity variation [Vinogradov 01], [Simovski 09], [Landau&Lifshits], Lorentz-Lorentz and Clausius-Mossotti equation [Clausius 78], [Mossotti 50], [Lorentz 10], [Ewald 21], and Bruggeman equation [Bruggeman 35] are no more valid. All these mentioned above theories have been elaborated in the framework of the static approximation, where electric and magnetic effects are separated. In the case of MM the main effect is in appearance of a magnetic response under the interaction of the metaatoms with the electric field, which is not included in the static approximation. An attempt to include dynamic corrections into the Lorenz-Lorentz expression, obtained in [Simovski 03], [Simovski 07] loses, as it was admitted by author, its physical meaning for the MM frequencies ranges [Simovski 09].

### 1.3 Frequency range of homogenization

It is methodologically important to distinguish between MM and other forms of compound structures like, for example, photonic crystals. In addition to the typical sizes (see Fig. 1) it is useful to introduce two characteristic wavelengths, namely:

1. Wavelength of the plane wave in the effective medium $\lambda_{eff} = \frac{2\pi}{\mathrm{Re}(k)}$ .
2. The resonance wavelength $\lambda_{res}$ of the internal resonances of the inclusions.

We are by definition interested in the wavelength region where the internal resonances of the inclusions exist $\lambda_{res} \sim \lambda_{eff}$ . The relation between the effective wavelength and the typical distance between inclusions $\lambda_{eff} > \frac{L_{inter}}{2}$ guaranties that the effective wavelength is safely longer than the wavelength of the Bragg resonances, otherwise the effect of mutual interference of the scattered waves dominates and the media is no more optically dense. Nevertheless, it is possible to create media where one of the typical sizes of the inclusions exceeds the effective wavelength significantly but the media can be homogenized anyway. For example, the media consisting of long and short wires can be homogenized and thus far can be considered as MM [Shvets 02], [Belov 03], [Silveirinha 06], [Severin 11].

The question about possible frequencies where the homogenization is possible has been considered in details in [Simovski 09] and [Simovski 11] (see also references therein). First, the effect of anisotropy which leads to different wave vectors for different propagation directions and which is typical for crystals, has to be taken into account. It was shown that the effect of anisotropy leads to two possible families of isofrequency contours, namely ellipsoids and hyperboloids, depending on the sign of the ratios $\frac{\varepsilon_{xx}}{\varepsilon_{zz}}$ and $\frac{\varepsilon_{yy}}{\varepsilon_{zz}}$ (here the consideration is carried out in the main axis of the intrinsic coordinate system of the crystal) [Sarychev 07], [Smith 03], [Belov 05]. From the other side, the effective parameters themself do not depend on the wave vector (spatial dispersion is negligible). In media with spatial dispersion the effective parameters depend on the wave vector considerably, therefore making the shape of the isofrequency contours arbitrary. Comparison of isofrequencies of MM obtained numerically with ellipsoids and hyperboloids makes it possible to identify frequency intervals, where the effective parameters do not depend on the wave vector. The conclusion in these papers is that the homogenization is possible only for the frequencies possessing isofrequency curves of ellipsoid or hyperboloid types in isofrequency contours – see, for example, Fig. 1 in [Simovski 09]. In other words, it was concluded that homogenization is possible only in case of absence of spatial dispersion. This conclusion is questionable. Keeping in mind, that the goal of the bulk homogenization is in finding a dispersion relation it has to be admitted, that this goal is fully achieved by all (not only by ellipsoid/hyperboloid) curves in the isofrequency contour. In other words, the fact that the dispersion relation (and the respective refractive index) depends on the direction does not contradict to the possibility of homogenization.

In case of periodic media there is a frequency range that corresponds to the absence of higher-order propagating Bloch modes. According to [Joannopoulis 95], [Rockstuhl 08], [Simovski 09], [Jackson 99], [Born 54], [Silveirinha 2011], this is the requirement of possibility of homogenization. This determination appeared to be questionable as well: the presence of several possible modes at the same frequency does not mean that the media could not be substituted by a homogeneous one with the same dispersion characteristics. The presence of several possible modes just mean that the propagating field will be presented as a sum of these modes with appropriate weighting coefficients; the question about relation between these coefficients is related to the problem of excitation of these modes, which in turn concerns the boundary condition problem, which is out of the scope of this paper.

Nevertheless, for the sake of simplicity and in order to fix the notations, it is assumed here that all characteristic sizes of the media (the sizes of the metaatoms and the distances between them) are smaller than one half of the effective wavelength in the media, which separates the problem of homogenization of MMs from the same problem for photonic crystal structures.

### 1.4. Different representations of material equation

The last equation in (4) is not averaged and describes the microscopic dynamics which is supposed to be substituted in $\langle \vec{j} \rangle$, $\langle \rho \rangle$ and averaged in order to get (5). Relations (5) in turn use information about microscopic dynamics as a function of microscopic fields which get averaged after substitution into equations for $\langle \vec{j} \rangle$, $\langle \rho \rangle$. In fact, there is only one model (a multipole model [Mazur 53]) where the averaging procedure for the $\langle \vec{j} \rangle$, $\langle \rho \rangle$ as functions of microscopic dynamics is performed rigorously, all other models do not even try to make this step; the last equation in (4) is usually left out completely, and the necessary equivalent information about charge dynamics is brought to the model phenomenologically. It has to be realized that this gap in the theory (inability to average rigorously the dynamic equation for particles in (4)) causes all the problems in further consideration and leads to all appearing contradictions and ambiguities.

In this paper we do not suggest any new ways of inclusion of this dynamic equation (the mentioned last equation in (4)) into consideration; instead, after this fact (ignorance of the dynamic equation from the averaging procedure) has been recognized, we try to create an approach to homogenization rigorously in terms of a logical chain with clearly recognized steps, assumptions, and approximations.

The system of equations (4), (5) is rather useless in practice until we find analytical expressions for (5). Nevertheless, even without finding of an analytical form for (5), the averaged MEs can be analysed and important conclusions can be made.

It is worth noticing that if we assume some analytical form for (5) (see, for example, [Menzel 10] and references herein) then the averaging problem is basically fixed (or, better to say, bypassed). System (4) becomes self-consistent and can be solved for the electric and magnetic fields $\vec{E}$, $\vec{B}$. Any further considerations (including introduction of $\vec{D}$, $\vec{H}$ in different representations, as well as the permittivity and permeability) in this case are no more required. Thus, in what follows we assume that there are no explicit forms of (5) and it is necessary to elaborate (5) further in order to find some reasonable analytical expressions for the averaged charge and current densities. It has to be emphasized as well that both these characteristics – the averaged charge and current densities – are measurable quantities and they do not change in math transformations; we will see below, that these characteristics keep their form in different ME representations.

First, following [Landau&Lifshits] we consider a volume with charges and fields (see again Fig. 1). The averaged charge in (5) can be represented through another function taking into account that the total charge of the considered volume is zero:

$$\int \langle \rho \rangle dV = 0 \quad (6)$$

It means that the averaged density of charges can be presented as a divergence of another unknown function $\vec{P}_{full}$ (see more details in [Landau&Lifshits] and [Vinogradov 01]):

$$\langle\rho\rangle = -\operatorname{div}\vec{P}_{full} \tag{7}$$

The function $\vec{P}_{full}$ is supposed to be zero outside the volume of integration in (6) [Landau&Lifshits]. In addition, this function is introduced with the accuracy of "rot" from any other arbitrary differentiable function $\vec{F}_1$:

$$\begin{cases} \langle\rho\rangle = -\operatorname{div}\vec{P}_{full} = -\operatorname{div}\left(\vec{P} + \operatorname{rot}\vec{F}_1\right) \\ \vec{P}_{full} = \vec{P} + \operatorname{rot}\vec{F}_1 \end{cases} \tag{8}$$

The averaged current is connected with the averaged charge density through the continuity relation [Bredov 85], which remains valid for the macroscopic representation:

$$\begin{cases} i\omega\langle\rho\rangle = \operatorname{div}\left\langle\vec{j}\right\rangle \\ \langle\rho\rangle = -\operatorname{div}\vec{P}_{full} \end{cases} \tag{9}$$

which gives:

$$\operatorname{div}\left(\left\langle\vec{j}\right\rangle + i\omega\vec{P}_{full}\right) = 0 \tag{10}$$

This means, that the averaged current can be introduced with the accuracy of "rot" of one more arbitrary function $\vec{F}_2$:

$$\left\langle\vec{j}\right\rangle = -i\omega\vec{P}_{full} = -i\omega\vec{P}_{full} + \operatorname{rot}\vec{F}_2 \tag{11}$$

or, taking into account (8):

$$\left\langle \vec{j} \right\rangle = -i\omega \vec{P}_{full} + \mathrm{rot}\,\vec{F}_2 = -i\omega \vec{P} + \mathrm{rot}\left(-i\omega \vec{F}_1 + \vec{F}_2\right) \tag{12}$$

It turns out that the material equations (3) can be written through one new function $\vec{P}$ with the accuracy of two more arbitrary functions $\vec{F}_1$ and $\vec{F}_2$:

$$\begin{cases} \left\langle \rho \right\rangle = -\mathrm{div}\left(\vec{P} + \mathrm{rot}\,\vec{F}_1\right) \\ \left\langle \vec{j} \right\rangle = -i\omega \vec{P} + \mathrm{rot}\left(-i\omega \vec{F}_1 + \vec{F}_2\right) \end{cases} \tag{13}$$

This approach assumes that the introduced function $\vec{P}_{full}$ is zero outside the volume of integration in (6), and, moreover, the function $\vec{P}_{full}$ does not depend on the chosen integration volume. Another approach has been proposed in [Vinogradov 99] and is called the scaling algorithm. The developed approach is based on a lemma proving that any field can be represented as a sum of three terms which are called “electric dipole”, “magnetic dipole”, and “electric quadrupole” moments (in the frequency domain):

$$J_i = -i\omega p_i + c e_{ijk} \frac{\partial m_k}{\partial x_j} - c \frac{\partial}{\partial x_k} \frac{\partial q_{ik}}{\partial t} = J_i^{(p)} + J_i^{(m)} + J_i^{(q)}$$

$$\begin{cases} m_i\left(x_j, J_k\right) = \dfrac{1}{2c} e_{ijk} x_j J_k \\ -i\omega q_{ij}\left(x_j, J_k\right) = \dfrac{1}{2c}\left(x_j J_i + x_i J_j\right) \\ -i\omega p_i\left(x_i, J_k\right) = -\left(x_j \dfrac{\partial J_k}{\partial x_k}\right) \end{cases} \tag{14}$$

This lemma leads to the possibility to represent, for example, the averaged current in the following form [Vinogradov 99]:

$$\left\langle \vec{j} \right\rangle = -i\omega\left(\vec{P} - \mathrm{div}\,Q\right) + c\,\mathrm{rot}\,\vec{M} \tag{15}$$

which basically repeats the second equation in (13). The equation for the averaged current $\left\langle \rho \right\rangle$ then becomes:

$$\langle\rho\rangle = -\operatorname{div}\left(\vec{P} - \operatorname{div} Q\right) \tag{16}$$

which is similar to the first equation in (13) if we assume that $\vec{P} \to \vec{P} - \operatorname{div} Q$. It is also rather straightforward to extend the approach (14) and include an analogy with the arbitrary functions $\vec{F}_1$ and $\vec{F}_2$. First, the function $\vec{M}$ in (15) appears to be the same as the function $\vec{F}_2$ (basically, $\vec{F}_2 = c\vec{M}$), and the function $\vec{F}_1$ can be included in (14) as extra terms for the magnetic (first equation in (14)) and dipole (last equation in (14)) terms:

$$J_i = -i\omega p_i + c e_{ijk} \frac{\partial m_k}{\partial x_j} - c \frac{\partial}{\partial x_k} \frac{\partial q_{ik}}{\partial t} = J_i^{(p)} + J_i^{(m)} + J_i^{(q)}$$

$$\begin{cases} m_i\left(x_j, J_k\right) = \dfrac{1}{2c} e_{ijk} x_j J_k + \dfrac{i\omega}{c} F_{1,i} \\ -i\omega q_{ij}\left(x_j, J_k\right) = \dfrac{1}{2c}\left(x_j J_i + x_i J_j\right) \\ -i\omega p_i\left(x_i, J_k\right) = -\left( x_j \dfrac{\partial J_k}{\partial x_k} \right) - e_{ijk} \dfrac{\partial F_{1,k}}{\partial x_j} \end{cases} \tag{17}$$

It is worth noting that equations (14) - (17) have been obtained without any additional assumptions about integration which have been used at the elaboration (7) from (6). The question about the physical meaning of the functions in (14) - (17) and (13) remains opened.

Now we come back to the consideration of (13). Both functions $\vec{F}_1$ and $\vec{F}_2$ are arbitrary and independent. This means that it is possible to consider different possibilities and to impose any additional requirements on them. There are different but countable number of choices for the possible representations of (13). The most general case is when both $\vec{F}_1$ and $\vec{F}_2$ are non-zero functions, namely:

$$\begin{cases} \vec{P}_{full} = \vec{P}_C = \vec{P} + \operatorname{rot} \vec{F}_1 \\ \vec{F}_2 = c * \vec{M}_C \end{cases} \tag{18}$$

which leads to the so called Casimir (subscript "C" stands for Casimir) form of material equations:

$$\begin{cases} \langle \rho \rangle = -\operatorname{div} \vec{P}_C \\ \langle \vec{j} \rangle = -i\omega \vec{P}_C + c \operatorname{rot} \vec{M}_C \end{cases} \qquad \begin{cases} \vec{D} = \vec{E} + 4\pi \vec{P}_C \\ \vec{H} = \vec{B} - 4\pi \vec{M}_C \end{cases} \tag{19}$$

In this case MEs include four functions $\vec{E},\ \vec{B}, \vec{D}, \vec{H}$ :

$$\begin{cases} \operatorname{rot} \vec{E} = \frac{i\omega}{c} \vec{B} \\ \operatorname{div} \vec{B} = 0 \\ \operatorname{div} \vec{D} = 0 \\ \operatorname{rot} \vec{H} = -\frac{i\omega}{c} \vec{D} \end{cases} \tag{20}$$

Note, that the case $\vec{F}_1 = 0$ and $\vec{F}_2 = c * \vec{M}_C$ leads to the same form (19), where the curl part of the full polarisability $\vec{P}_{full}$ (see (8)) is excluded (the physical meaning of this part – presence of anapoles [Zeldovich 57] - will be considered below).

Alternatively to (18), we can set:

$$\begin{cases} \vec{P}_{full} = \vec{P}_{LL} = \vec{P} + \operatorname{rot} \vec{F}_1 \\ \vec{F}_2 = 0 \end{cases} \tag{21}$$

which leads according to (13) to the so called Landau&Lifshitz (subscript “L&L” stands for Landau&Lifshitz) [Mandelshtam 57] form of material equations:

$$\begin{cases} \langle \rho \rangle = -\operatorname{div} \vec{P}_{LL} \\ \langle \vec{j} \rangle = -i\omega \vec{P}_{LL} \end{cases} \qquad \begin{cases} \vec{D} = \vec{E} + 4\pi \vec{P}_{LL} \\ \vec{B} = \vec{B} \end{cases} \tag{22}$$

In this case MEs contain three functions $\vec{E},\ \vec{B}, \vec{D}$:

$$\begin{cases} \mathrm{rot}\,\vec{E} = \dfrac{i\omega}{c}\vec{B} \\ \mathrm{div}\,\vec{B} = 0 \\ \mathrm{div}\,\vec{D} = 0 \\ \mathrm{rot}\,\vec{B} = -\dfrac{i\omega}{c}\vec{D} \end{cases} \tag{23}$$

Note that the form (22) does not assume that the averaged current $\langle\vec{j}\rangle$ does not contain any curl part – this part is included in $\langle\vec{j}\rangle$ through $\vec{F}_1$ (see (21)). The main difference between "C" and "L&L" representation is in the absence in the latter any stationary (not proportional to $\omega$) part of the curl part of $\langle\vec{j}\rangle$, described by $\vec{M}_C$. In case of the absence of the stationary magnetization both representations have to be equivalent.

Finally, we assume that the full polarisability $\vec{P}_{full}$ contains only the curl part, namely:

$$\begin{cases} \vec{P}_{full} = \mathrm{rot}\,\vec{F}_1 \\ \vec{F}_2 = c * \vec{M}_C \end{cases} \tag{24}$$

which leads, according to (13), to the case which we call here Anapole ( subscript "A" stands for Anapole) form of material equations:

$$\begin{cases} \langle\rho\rangle = 0 \\ \langle\vec{j}\rangle = -i\omega\,\mathrm{rot}\,\vec{F}_1 + c\,\mathrm{rot}\,\vec{M}_C = c\,\mathrm{rot}\,\vec{M}_A \end{cases} \qquad \begin{cases} \vec{D} = \vec{E} + 4\pi\,\mathrm{rot}\,\vec{F}_1 \\ \vec{H} = \vec{B} - 4\pi\vec{M}_A \end{cases} \tag{25}$$

In this case the system of MEs contains three functions $\vec{E}$, $\vec{B}$, $\vec{H}$ and reads:

$$\begin{cases} \mathrm{rot}\,\vec{E} = \dfrac{i\omega}{c}\vec{B} \\ \mathrm{div}\,\vec{B} = 0 \\ \mathrm{div}\,\vec{E} = 0 \\ \mathrm{rot}\,\vec{H} = -\dfrac{i\omega}{c}\vec{E} \end{cases} \tag{26}$$

We stress that this set of equations can be used only in very special cases where the averaged charge density is zero. Thus, in general the anapole form cannot be used *instead* of the Casimir or Landau-Lifshits forms.

The physical object corresponding to such representation is an anapole [Zeldovich 57], [Vinogradov 01], which is now of great interest in connection with potential possibility of design of such structures at nanoscales for optical wavelength region application [Kaelberer 10]. It is seen, that the presence of anapoles is responsible for the function $\vec{F}_1$, and fixes the functions $\langle\rho\rangle, \langle\vec{j}\rangle$ in general representation (13).

*One can conclude that the anapoles play a fundamental role in electrodynamics, particularly in the averaging procedure for MEs, unambiguously fixing representation for the polarizability and magnetization of the averaged media. The anapoles exhibit themselves as the zero averaged charge objects providing nevertheless nonzero magnetization* $\vec{M}$ *and hence playing a role of "dark matter" in electrodynamics.* The principal difference with quadrupoles is in the fact that these objects couple to external fields and create secondary fields outside. In contrast, in the case of an ideal anapole, all its fields are concentrated inside the anapole object.

Physical interpretation of the three mentioned above representations can be done based on the types of atoms/molecules (or metaatoms/metamolecules) which the considered media consist of. In the most general case (multipoles, dynamic and stationary magnetization, and anapoles) "C" form is preferable. In the case of absence of stationary magnetisation (but presence of all others) the "C" and "L&L" representations have to be equivalent. In case of absence of the non curl part of magnetization (presence of only anapoles and maybe stationary magnetization) the "A" form is appropriate.

*It is important to realize that there are no other choices for the material equations. Any homogenization model has to start from the statement in which representation it will be developed; arbitrary mixing between several representations is not acceptable, as it will be seen below.*

The possible representations are summarized in Table 1.

| **Casimir form ("C" form)** $\vec{E},\ \vec{B}, \vec{D}, \vec{H}$ | **Landau & Lifshitz form ("LL" form)** $\vec{E},\ \vec{B}, \vec{D}, \cancel{\vec{H}}$ | **Anapole form ("A" form)** $\vec{E},\ \vec{B}, \cancel{\vec{D}}, \vec{H}$ |
|---|---|---|
| $\begin{cases}\langle\rho\rangle = -\operatorname{div}\vec{P}_C \\ \langle\vec{j}\rangle = -i\omega\vec{P}_C + c\operatorname{rot}\vec{M}_C\end{cases}$ | $\begin{cases}\langle\rho\rangle = -\operatorname{div}\vec{P}_{LL} \\ \langle\vec{j}\rangle = -i\omega\vec{P}_{LL}\end{cases}$ | $\begin{cases}\langle\rho\rangle = 0 \\ \langle\vec{j}\rangle = c\operatorname{rot}\vec{M}_A\end{cases}$ |
| $\begin{cases}\vec{D} = \vec{E} + 4\pi\vec{P}_C \\ \vec{H} = \vec{B} - 4\pi\vec{M}_C\end{cases}$ | $\begin{cases}\vec{D} = \vec{E} + 4\pi\vec{P}_{LL} \\ \vec{H} = \vec{B}\end{cases}$ | $\begin{cases}\vec{D} = \vec{E} \\ \vec{H} = \vec{B} - 4\pi\vec{M}_A\end{cases}$ |
| $\begin{cases}\operatorname{rot}\vec{E} = \frac{i\omega}{c}\vec{B} \\ \operatorname{div}\vec{B} = 0 \\ \operatorname{div}\vec{D} = 0 \\ \operatorname{rot}\vec{H} = -\frac{i\omega}{c}\vec{D}\end{cases}$ | $\begin{cases}\operatorname{rot}\vec{E} = \frac{i\omega}{c}\vec{B} \\ \operatorname{div}\vec{B} = 0 \\ \operatorname{div}\vec{D} = 0 \\ \operatorname{rot}\vec{B} = -\frac{i\omega}{c}\vec{D}\end{cases}$ | $\begin{cases}\operatorname{rot}\vec{E} = \frac{i\omega}{c}\vec{B} \\ \operatorname{div}\vec{B} = 0 \\ \operatorname{div}\vec{E} = 0 \\ \operatorname{rot}\vec{H} = -\frac{i\omega}{c}\vec{E}\end{cases}$ |

**Table 1: Possible forms of representation of material equations and the respective forms of the macroscopic MEs.**

## 1.5. Serdyukov-Fedorov transformation between different representations

The Serdyukov-Fedorov transformations (SFT) are relations between two sets of four vectors $\vec{E},\ \vec{B}, \vec{D}, \vec{H}$ and $\vec{E}',\ \vec{B}', \vec{D}', \vec{H}'$, where both sets satisfy Maxwell's equations. The SFT are usually written in the following form (for "C" representation):

$$\begin{cases} \operatorname{rot}\vec{E} = \frac{i\omega}{c}\vec{B} \\ \operatorname{div}\vec{B} = 0 \\ \operatorname{div}\vec{D} = 0 \\ \operatorname{rot}\vec{H} = -\frac{i\omega}{c}\vec{D} \end{cases} \qquad \begin{cases} \vec{B} = \vec{B}' + \operatorname{rot}\vec{T}_1 \\ \vec{E} = \vec{E}' + \frac{i\omega}{c}\vec{T}_1 \\ \vec{D} = \vec{D}' + \operatorname{rot}\vec{T}_2 \\ \vec{H} = \vec{H}' - \frac{i\omega}{c}\vec{T}_2 \end{cases} \tag{27}$$

The SFT are composed from two parts - “field” transformations for $\vec{E}$, $\vec{B}$ and “material” transformations for $\vec{D}$, $\vec{H}$ ). Below SFT are used to find relations between the three forms of material relations.

First of all, let us consider the system of averaged MEs in the following form:

$$\begin{cases} \operatorname{rot}\vec{E} = \frac{i\omega}{c}\vec{B} \\ \operatorname{div}\vec{B} = 0 \\ \operatorname{div}\vec{E} = 4\pi\langle\rho\rangle \\ \operatorname{rot}\vec{B} = -\frac{i\omega}{c}\vec{E} + \frac{4\pi}{c}\langle\vec{j}\rangle \end{cases} \tag{28}$$

It is easy to see that the first two equations are invariant under the following transformation:

$$\begin{cases} \vec{B} = \vec{B}' + \operatorname{rot}\vec{T}_1 \\ \vec{E} = \vec{E}' + \frac{i\omega}{c}\vec{T}_1 \end{cases} \tag{29}$$

The new primed fields satisfy the same form of the first two MEs (28). It corresponds exactly to the SFT in parts of the field transformations in (27).

Substituting (29) into the last two equations of (28), it is easy to find relations between new and old averaged charge density and current:

$$\begin{cases} \langle\rho\rangle = \langle\rho\rangle' + \frac{i\omega}{4\pi c}\operatorname{div}\vec{T}_1 \\ \langle\vec{j}\rangle = \langle\vec{j}\rangle' + \frac{c}{4\pi}\operatorname{rot}\operatorname{rot}\vec{T}_1 - \frac{\omega^2}{4\pi c}\vec{T}_1 \end{cases} \tag{30}$$

Transformations (29) and (30) give relations between fields and charge and current densities so that the functional form of the MEs remains the same. It should be emphasized that at this stage absolutely no assumptions about possible distributions of the charge and current densities have been made.

Physical interpretation of the found transformations is pretty obvious: they just link two possible solutions of MEs. For example, from a known (non-primed) solution of MEs one can find new (primed) one if we know the connection between the non-primed and primed averaged charge and current densities (30). A more important consequence of (30) is that in case of the field transformation (29) it is necessary to transform charge and current densities (30), if we require that the new primed values satisfy the MEs.

Previously it has been found that the material equations (5) can be written through two new functions $\vec{P}$ and $\vec{M}$, which for the most general "C" representation takes the form (see (19)):

$$\begin{cases} \langle \rho \rangle = -\operatorname{div} \vec{P} \\ \langle \vec{j} \rangle = -i\omega \vec{P} + c \operatorname{rot} \vec{M} \end{cases} \tag{31}$$

From (31) one can immediately obtain that in case of transformation (30) functions $\vec{P}$ and $\vec{M}$ have to be transformed as well:

$$\begin{cases} \vec{P} = \vec{P}' - \dfrac{i\omega}{4\pi c} \vec{T}_1 \\ \vec{M} = \vec{M}' + \dfrac{1}{4\pi} \operatorname{rot} \vec{T}_1 \end{cases} \tag{32}$$

Substituting (31) into MEs we obtain:

$$\begin{cases} \operatorname{rot} \vec{E} = \dfrac{i\omega}{c} \vec{B} \\ \operatorname{div} \vec{B} = 0 \\ \operatorname{div}\left( \vec{E} + 4\pi \vec{P} \right) = 0 \\ \operatorname{rot}\left( \vec{B} - 4\pi \vec{M} \right) = -\dfrac{i\omega}{c} \left( \vec{E} + 4\pi \vec{P} \right) \end{cases} \tag{33}$$

It is clear that the last two equations allow similar transformation with the use of one more arbitrary function $\vec{T}_2$, as in the second set of (27). Namely,

$$\begin{cases} \vec{P} = \vec{P}' - \dfrac{i\omega}{4\pi c}\vec{T}_1 + \mathrm{rot}\,\vec{T}_2 \\ \vec{M} = \vec{M}' + \dfrac{1}{4\pi}\mathrm{rot}\,\vec{T}_1 + \dfrac{i\omega}{c}\vec{T}_2 \end{cases} \tag{34}$$

which also does not change the form of the MEs.

It should be emphasized that function $\vec{T}_2$ does not appear in the expressions for transformation of averaged charge and current densities; in other words, measurable values are not changed under transformation (34).

As discussed above, the physical meaning of the transformations (29), (30) is in just an algebraic link between two possible solutions of MEs: if the averaged charge and current densities are transformed according to (30), then the new fields can be obtained without the necessity to solve MEs, but with the use of (29). The physical interpretation of (34) with $\vec{T}_1 = 0$ is different. This transformation changes neither fields nor averaged densities of charges and currents, but rather redistributes the representations of them between $\vec{P}$ and $\vec{M}$. In other words, this transformation describes the same physical situation by different representations, e.g. by different pairs of $\vec{P}, \vec{M}$. It is clear (and it will be used below) that this corresponds to transformations between different representations of MEs ("C", "L&L", and "A" forms).

Alternatively, in (32) one can introduce transformation not for $\vec{P}, \vec{M}$, but for combinations $\vec{E} + 4\pi\vec{P},\ \vec{B} - 4\pi\vec{M}$, namely:

$$\begin{cases} \vec{E} + 4\pi\vec{P} = \vec{E}' + 4\pi\vec{P}' + \mathrm{rot}\,\vec{T}_2 \\ \vec{B} - 4\pi\vec{M} = \vec{B}' - 4\pi\vec{M}' - \dfrac{i\omega}{c}\vec{T}_2 \end{cases} \tag{35}$$

which is equivalent to the last two equations in (27):

$$\begin{cases} \vec{D} = \vec{D}' + \mathrm{rot}\,\vec{T}_2 \\ \vec{H} = \vec{H}' - \dfrac{i\omega}{c}\vec{T}_2 \end{cases} \tag{36}$$

Combining all elaborated expressions, one can finally obtain, that the transformations

$$\begin{cases} \operatorname{rot}\vec{E} = \frac{i\omega}{c}\vec{B} \\ \operatorname{div}\vec{B} = 0 \\ \operatorname{div}\left(\vec{E} + 4\pi\vec{P}\right) = 0 \\ \operatorname{rot}\left(\vec{B} - 4\pi\vec{M}\right) = -\frac{i\omega}{c}\left(\vec{E} + 4\pi\vec{P}\right) \\ \langle\rho\rangle = -\operatorname{div}\vec{P} \\ \langle\vec{j}\rangle = -i\omega\vec{P} + c\operatorname{rot}\vec{M} \end{cases} \qquad \begin{cases} \vec{B} = \vec{B}' + \operatorname{rot}\vec{T}_1 \\ \vec{E} = \vec{E}' + \frac{i\omega}{c}\vec{T}_1 \\ \vec{P} = \vec{P}' - \frac{i\omega}{4\pi c}\vec{T}_1 + \operatorname{rot}\vec{T}_2 \\ \vec{M} = \vec{M}' + \frac{1}{4\pi}\operatorname{rot}\vec{T}_1 + \frac{i\omega}{c}\vec{T}_2 \\ \langle\rho\rangle = \langle\rho\rangle' + \frac{i\omega}{4\pi c}\operatorname{div}\vec{T}_1 \\ \langle\vec{j}\rangle = \langle\vec{j}\rangle' + \frac{c}{4\pi}\operatorname{rot}\operatorname{rot}\vec{T}_1 - \frac{\omega^2}{4\pi c}\vec{T}_1 \end{cases} \qquad (37)$$

are equivalent to the SFT (27). The difference between SFT in form (27) and (37) is in that in the developed here approach functions $\vec{P},\vec{M}$ are used instead of $\vec{D},\vec{H}$. The latter variant appears to be convenient due to the fact that in this case transformation for the fields $\vec{E},\vec{B}$ is independent from the transformation for $\vec{D},\vec{H}$. Nevertheless, it is important to remind that in order to come to the representation using vectors $\vec{D},\vec{H}$, it is necessary first to introduce $\vec{P},\vec{M}$, after that regroup the respective terms in MEs, and then introduce vectors $\vec{D},\vec{H}$. It is also possible to postulate the macroscopic MEs directly in form (27), but in this case the logical transition between microscopic and macroscopic MEs is lost.

*Concluding this part, we can state that the SFT provide transformations between two different realizable physical situations (* $\vec{T}_1 \neq 0, \vec{T}_2 = 0$ *) or between two representations of the same physical situation (* $\vec{T}_1 = 0, \vec{T}_2 \neq 0$ *).*

### 1.6. Transformation between different representations

Let us consider the relation between different representations of MEs, and start from the “C” form (19). It is known that the “C” form is invariant with respect to the SFT (27), (37). Here the SFT will be applied (following [Vinogradov 01]) and possible conclusions which can be made based on the application of the SFT will be considered.

#### “C” to “L&L” transformation

Let us start from “C” to “L&L” transformation. For the material equations in “C” (not primed) and “L&L” (primed) forms one can respectively write:

$$\begin{cases}\vec{B}_C = \vec{B}'_{LL} + \operatorname{rot}\vec{T}_1 \\ \vec{E}_C = \vec{E}'_{LL} + \dfrac{i\omega}{c}\vec{T}_1\end{cases} \qquad \begin{cases}\vec{P}_C = \vec{P}'_{LL} - \dfrac{i\omega}{4\pi c}\vec{T}_1 + \operatorname{rot}\vec{T}_2 \\ \vec{M}_C = \vec{M}'_{LL} + \dfrac{i\omega}{c}\vec{T}_2 + \dfrac{\operatorname{rot}\vec{T}_1}{4\pi} \\ \langle\rho\rangle_C = \langle\rho\rangle'_{LL} + \dfrac{i\omega}{4\pi c}\operatorname{div}\vec{T}_1 \\ \langle\vec{j}\rangle_C = \langle\vec{j}\rangle'_{LL} + \dfrac{c}{4\pi}\operatorname{rot}\operatorname{rot}\vec{T}_1 - \dfrac{\omega^2}{4\pi c}\vec{T}_1\end{cases} \tag{38}$$

In order to get the MEs for the new fields in “L&L” form in the primed system we have to require that:

$$\vec{M}'_{LL} = 0 \tag{39}$$

which leads to:

$$\vec{T}_2 = \frac{ic}{\omega}\left(\operatorname{rot}\vec{T}_1 - \vec{M}_C\right) \tag{40}$$

Substituting the last equation into (38), we finally have:

$$\begin{cases}\vec{B}_C = \vec{B}'_{LL} + \operatorname{rot}\vec{T}_1 \\ \vec{E}_C = \vec{E}'_{LL} + \dfrac{i\omega}{c}\vec{T}_1\end{cases} \qquad \begin{cases}\vec{P}'_{LL} = \vec{P}_C + \dfrac{i\omega}{4\pi c}\vec{T}_1 + \dfrac{ic}{\omega}\operatorname{rot}\left(\operatorname{rot}\vec{T}_1 - \vec{M}_C\right) \\ \vec{M}'_{LL} = 0 \\ \langle\rho\rangle_C = \langle\rho\rangle'_{LL} + \dfrac{i\omega}{4\pi c}\operatorname{div}\vec{T}_1 \\ \langle\vec{j}\rangle_C = \langle\vec{j}\rangle'_{LL} + \dfrac{c}{4\pi}\operatorname{rot}\operatorname{rot}\vec{T}_1 - \dfrac{\omega^2}{4\pi c}\vec{T}_1\end{cases} \tag{41}$$

which gives us the MEs in form of “L&L”. If, in addition, we require that the electric and magnetic fields remain the same for both representations (which is reasonable, because both fields are assumed to be physically measurable), we obtain by setting $\vec{T}_1$ to zero:

$$\begin{cases} \vec{B}_C = \vec{B}'_{LL} \\ \vec{E}_C = \vec{E}'_{LL} \end{cases} \qquad \begin{cases} \vec{P}'_{LL} = \vec{P}_C - \frac{ic}{\omega}\operatorname{rot}\vec{M}_C \\ \vec{M}'_{LL} = 0 \\ \langle\rho\rangle_C = \langle\rho\rangle'_{LL} \\ \langle\vec{j}\rangle_C = \langle\vec{j}\rangle'_{LL} \end{cases} \tag{42}$$

We see that starting from "C" representation, we can unambiguously reduce the MEs to the "L&L" form. It is important to emphasize, that in general ($\vec{T}_1 \neq 0$) both electric and magnetic fields are transformed and lose their initial physical meanings. The requirement of keeping the electric and magnetic fields the same in both representations is an additional one with respect to the SFT.

### "L&L" to "C" transformation

Let us consider the inverse transformation ("L&L" to "C" representation), namely we start from system (22) and write the SFT in this case:

$$\begin{cases} \vec{B}_{LL} = \vec{B}'_C + \operatorname{rot}\vec{T}_1 \\ \vec{E}_{LL} = \vec{E}'_C + \frac{i\omega}{c}\vec{T}_1 \end{cases} \qquad \begin{cases} \vec{P}_{LL} = \vec{P}'_C - \frac{i\omega}{4\pi c}\vec{T}_1 + \operatorname{rot}\vec{T}_2 \\ 0 = \vec{M}'_C + \frac{i\omega}{c}\vec{T}_2 + \frac{\operatorname{rot}\vec{T}_1}{4\pi} \\ \langle\rho\rangle_{LL} = \langle\rho\rangle'_C + \frac{i\omega}{4\pi c}\operatorname{div}\vec{T}_1 \\ \langle\vec{j}\rangle_{LL} = \langle\vec{j}\rangle'_C + \frac{c}{4\pi}\operatorname{rot}\operatorname{rot}\vec{T}_1 - \frac{\omega^2}{4\pi c}\vec{T}_1 \end{cases} \tag{43}$$

It is seen, that in general the functions $\vec{T}_1$, $\vec{T}_2$ cannot be unambiguously found.

As like as for direct ("C" to "L&L") transformation, we require that the field in both representations remain the same $\vec{T}_1 = 0$:

$$\begin{cases} \vec{B}_{LL} = \vec{B}'_C \\ \vec{E}_{LL} = \vec{E}'_C \end{cases} \qquad \begin{cases} \vec{P}'_C = \vec{P}_{LL} - \operatorname{rot}\vec{T}_2 \\ \vec{M}'_C = -\frac{i\omega}{c}\vec{T}_2 \\ \langle\rho\rangle_{LL} = \langle\rho\rangle'_C \\ \langle\vec{j}\rangle_{LL} = \langle\vec{j}\rangle'_C \end{cases} \tag{44}$$

It is seen that the function $\vec{T}_2$ cannot be unambiguously determined, and the reverse "C" to "L&L" transformation is in general undetermined as well. This is because splitting of the total polarization current into two parts cannot be made uniquely and some additional physical requirements are needed to define, as is usually done, the parts corresponding to electric polarization density and magnetization current.

Let us emphasize again, that starting from the "C" form it is possible to arrive to the "L&L" form, but starting from the "L&L" form it is impossible to reduce MEs to the "C" form using the SFT unambiguously. There are unlimited number of "C" forms which correspond to the same "L&L" form.

**"C" to "A" transformation**

Let us now consider the "C" to "A" transformation. In order to get the MEs for the new fields in "A" form (25) in the primed system we have to require that

$$\langle \rho \rangle_A = 0 \tag{45}$$

Writing the SFT for this case, we have:

$$\begin{cases} \vec{B}_C = \vec{B}'_A + \operatorname{rot}\vec{T}_1 \\ \vec{E}_C = \vec{E}'_A + \dfrac{i\omega}{c}\vec{T}_1 \end{cases} \qquad \begin{cases} \vec{P}_C = \vec{P}'_A - \dfrac{i\omega}{4\pi c}\vec{T}_1 + \operatorname{rot}\vec{T}_2 \\ \vec{M}_C = \vec{M}'_A + \dfrac{i\omega}{c}\vec{T}_2 + \dfrac{\operatorname{rot}\vec{T}_1}{4\pi} \\ \langle \rho \rangle_C = \dfrac{i\omega}{4\pi c}\operatorname{div}\vec{T}_1 \\ \langle \vec{j} \rangle_C = \langle \vec{j} \rangle'_A + \dfrac{c}{4\pi}\operatorname{rot}\operatorname{rot}\vec{T}_1 - \dfrac{\omega^2}{4\pi c}\vec{T}_1 \end{cases} \tag{46}$$

Substituting $\langle \rho \rangle_C = -\operatorname{div}\vec{P}_C$ into the third equation in (46) we obtain $\operatorname{div}\left(\vec{T}_1 - \dfrac{i4\pi c}{\omega}\vec{P}_C\right) = 0$, $\vec{P}_C = -\dfrac{i\omega}{4\pi c}\vec{T}_1 + \operatorname{rot}\vec{T}_3$ , $\vec{T}_3$ is arbitrary, and finally:

$$\begin{cases} \vec{B}_C = \vec{B}'_A + \operatorname{rot}\vec{T}_1 \\ \vec{E}_C = \vec{E}'_A + \dfrac{i\omega}{c}\vec{T}_1 \end{cases} \qquad \begin{cases} \vec{P}'_A = -\operatorname{rot}\left(\vec{T}_2 + \vec{T}_3\right) \\ \vec{M}'_A = \vec{M}_C + \dfrac{i\omega}{c}\vec{T}_2 - \dfrac{ic}{\omega}\operatorname{rot}\vec{P}_C \\ \left\langle \vec{j} \right\rangle_C = \left\langle \vec{j} \right\rangle'_A + \dfrac{ic^2}{\omega}\operatorname{rot}\operatorname{rot}\vec{P}_C - i\omega\vec{P}_C \end{cases} \tag{47}$$

and $\vec{P}'_A$ and $\vec{M}'_A$ are not unambiguously determined. In addition, if we require the same values for the electric and magnetic fields in both representations, we get $\langle \rho \rangle_C = 0$ (see (46)) which in general does not take place. One can conclude that the "C" to "A" transformation in general cannot be performed. Let us stress that this reflects the fact that the anapole form can be used only if the average charge density is zero, which is of course generally not the case.

**"A" to "C" transformation**

In this case the SFT has the following form:

$$\begin{cases} \vec{B}_A = \vec{B}'_C + \operatorname{rot}\vec{T}_1 \\ \vec{E}_A = \vec{E}'_C + \dfrac{i\omega}{c}\vec{T}_1 \end{cases} \qquad \begin{cases} \vec{P}_A = \vec{P}'_C - \dfrac{i\omega}{4\pi c}\vec{T}_1 + \operatorname{rot}\vec{T}_2 \\ \vec{M}_A = \vec{M}'_C + \dfrac{i\omega}{c}\vec{T}_2 + \dfrac{\operatorname{rot}\vec{T}_1}{4\pi} \\ \langle \rho \rangle'_C = -\dfrac{i\omega}{4\pi c}\operatorname{div}\vec{T}_1 \\ \left\langle \vec{j} \right\rangle_A = \left\langle \vec{j} \right\rangle'_C + \dfrac{c}{4\pi}\operatorname{rot}\operatorname{rot}\vec{T}_1 - \dfrac{\omega^2}{4\pi c}\vec{T}_1 \end{cases} \tag{48}$$

remembering $\langle \rho \rangle'_C = -\operatorname{div}\vec{P}'_C$ and consequently $\operatorname{div}\left(\vec{T}_1`+i\dfrac{4\pi c}{\omega}\vec{P}'_C\right)=0$, $\vec{P}'_C = \dfrac{i\omega}{4\pi c}\vec{T}_1 + \operatorname{rot}\vec{T}_3$, $\vec{T}_3$ is arbitrary, we get:

$$\begin{cases} \vec{B}_A = \vec{B}'_C + \operatorname{rot}\vec{T}_1 \\ \vec{E}_A = \vec{E}'_C + \dfrac{i\omega}{c}\vec{T}_1 \end{cases} \qquad \begin{cases} \vec{P}_A = \operatorname{rot}\left(\vec{T}_2 + \vec{T}_3\right) \\ \vec{M}'_C = \vec{M}_A + \dfrac{i\omega}{c}\vec{T}_2 + i\dfrac{c}{\omega}\operatorname{rot}\vec{P}'_C \\ \left\langle \vec{j} \right\rangle'_C = \left\langle \vec{j} \right\rangle_A + i\dfrac{c^2}{\omega}\operatorname{rot}\operatorname{rot}\vec{P}'_C - i\omega\vec{P}'_C \end{cases} \tag{49}$$

It is seen, that in this case $\vec{P}'_C$ is not determined unambiguously. In case of $\vec{T}_1 = 0$ (equivalence of the fields in both forms) we have $\vec{P}'_C = \mathrm{rot}\vec{T}_3$, both charge and current densities unchanged (as it should be) $\langle\rho\rangle'_C = 0$ and $\langle\vec{\mathrm{j}}\rangle'_C = \langle\vec{\mathrm{j}}\rangle_A$; the only function which has been changed is the magnetization $\vec{M}'_C = \vec{M}_A + \frac{i\omega}{c}\vec{T}_2 + i\frac{c}{\omega}\mathrm{rot}\,\mathrm{rot}\vec{T}_3$. $\vec{P}'_C$ and $\vec{M}'_C$ remain unfixed which means that "A" to "C" transformation cannot be unambiguously determined.

**"L&L" to "A" transformation**

In this case the SFT reads (taking into account from the beginning, that $\langle\rho\rangle_{LL} = -\mathrm{div}\vec{P}_{LL}$ and $\mathrm{div}\left(\vec{T}_1 - i\frac{4\pi c}{\omega}\vec{P}_{LL}\right) = 0$:

$$\begin{cases} \vec{B}_{LL} = \vec{B}'_A + \mathrm{rot}\vec{T}_1 \\ \vec{E}_{LL} = \vec{E}'_A + \frac{i\omega}{c}\vec{T}_1 \end{cases} \qquad \begin{cases} \vec{P}'_A = -\mathrm{rot}\left(\vec{T}_2 + \vec{T}_3\right) \\ \vec{M}'_A = -\frac{i\omega}{c}\vec{T}_2 - \frac{\mathrm{rot}\vec{T}_1}{4\pi} \\ \langle\rho\rangle_{LL} = \frac{i\omega}{4\pi c}\mathrm{div}\vec{T}_1 \\ \langle\vec{\mathrm{j}}\rangle'_A = \langle\vec{\mathrm{j}}\rangle_{LL} - \frac{c}{4\pi}\mathrm{rot}\,\mathrm{rot}\vec{T}_1 + \frac{\omega^2}{4\pi c}\vec{T}_1 \end{cases} \qquad (50)$$

The transformation, similarly to the case "C" to "A", in general is not determined due to the fact that in general $\langle\rho\rangle_{LL} \neq 0$, but in case of the same fields in both representations it would be required by the fact that $\vec{T}_1 = 0$. Mathematically this is clearly seen from the first equation in (50): in order to "move" the material response from the effective polarization vector in the constitutive relation for D (L&L form) into an effective magnetization vector in the constitutive relation for B ("A" form), the polarization vector must be represented as a curl of a certain vector. This obviously implies that its divergence is zero, that is, the charge density must be zero to allow such transformation.

**"A" to "L&L" transformation**

The reverse transformation follows the equations (again, the equation $\langle\rho\rangle'_{LL} = -\mathrm{div}\vec{P}'_{LL}$ is already taken into account):

$$\begin{cases} \vec{B}_A = \vec{B}'_{LL} + \mathrm{rot}\,\vec{T}_1 \\ \vec{E}_A = \vec{E}'_{LL} + \dfrac{i\omega}{c}\vec{T}_1 \end{cases} \qquad \begin{cases} \vec{P}_A = \mathrm{rot}\left(\vec{T}_2 + \vec{T}_3\right) \\ \vec{M}_A = \dfrac{i\omega}{c}\vec{T}_2 + \dfrac{1}{4\pi}\mathrm{rot}\,\vec{T}_1 \\ \langle\rho\rangle'_{LL} = -\dfrac{i\omega}{4\pi c}\mathrm{div}\,\vec{T}_1 \\ \langle\vec{\mathrm{j}}\rangle'_{LL} = \langle\vec{\mathrm{j}}\rangle_A - \dfrac{c}{4\pi}\mathrm{rot}\,\mathrm{rot}\,\vec{T}_1 + \dfrac{\omega^2}{4\pi c}\vec{T}_1 \end{cases} \tag{51}$$

In this case the transformation cannot be done at all: the “L&L” polarizability is excluded from consideration, and magnetization is zero by definition of the “L&L” representation.
The mutual transformations between different representations are presented in Fig. 4.

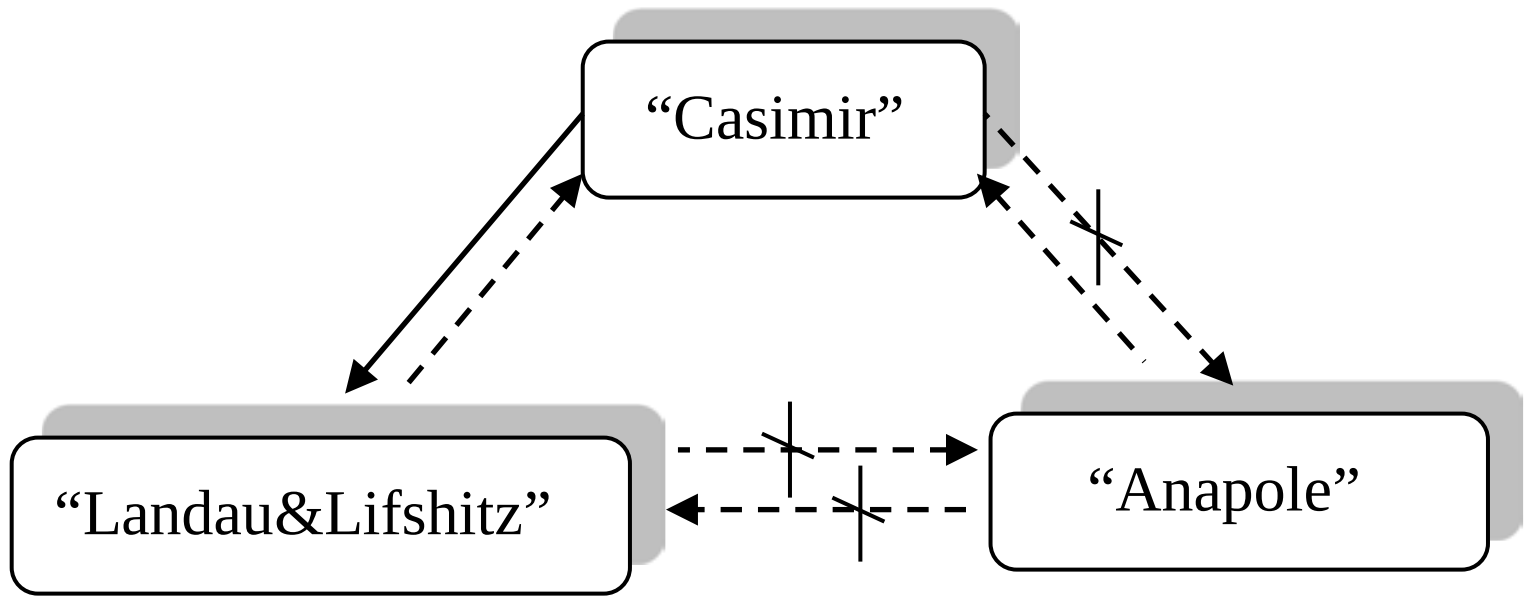

**Fig. 4: Possibility of mutual transformations between different representations. Crossed dashed lines between “Casimir” and “Anapole” and “Landau&Lifshitz” and “Anapole” show impossible transformations, multiple dashed lines between “L&L” and “C” show not unique transformations.**

From Fig. 4 one can see that the “L&L” form occupies a special place in the elaborated hierarchy – the other two forms can be reduced to the “L&L” form, while the “L&L” form itself cannot be transformed to the other unambiguously. Another specific position is occupied by the “A” representation: this representation is valid only if the averaged charge density is identically zero and it cannot be achieved in general from two other ones.

## 1.7. Conclusions of Part 1

All the considerations above did not answer the question “How to get the unknown functions for the polarizability (”L&L” form) or polarization and magnetization (“C” form) starting from the microscopic picture?”

The main problem is to develop a model, which would give us a recipe to find analytical expressions for $\vec{P}$ and $\vec{M}$ as functions of the averaged fields – it has to be also pointed out, that the expressions have to be presented as functions of the averaged (macroscopic), and not the microscopic fields; only in this case we can formulate the MEs as a self-consistent system. Nevertheless, it is important to realize, that whatever model is developed, it can be presented only in "C", "L&L", or "A" form with the respective consequences, described above.

The conclusions for the presented above chapter are:

1. In this chapter the main goal of this work has been formulated in form of defining the functional dependence (5) or, equivalently, (19), (22), (25). The goal has been formulated based on the microscopic Maxwell equations in form of fields (4) as a starting point for all considerations.
2. The frequency range where the homogenization procedure can be performed has been determined.
3. The macroscopic approach to the homogenization has been developed resulting in three possible representations of ME – "C", "L&L", and "A" forms.
4. The Serdyukov-Fedorov transformations (SFT) have been reformulated and used to establish relationships between the three ME representations.

## 2. Phenomenological vs multipole models

In order to further develop the homogenization procedure, we should find an analytical form for the functions $\vec{P}$ and $\vec{M}$ in case of "C" representation, $\vec{P}$ in case of "L&L" representation, or $\vec{M}$ in case of "A" representation. There are two principal ways to do it: introduce a functional form following the phenomenological approach, or try to create a microscopic model which would result finally in the required forms for $\vec{P}$ and $\vec{M}$. Below, both ways will be considered.

### 2.1. Phenomenological model

#### 2.1.1 "L&L" representation

***Case of strong spatial dispersion***

There is a commonly accepted integral form of $\vec{P}_{LL}$ , which can be written according to the causality principle (which imposes limitations on the frequency dispersion form) and assuming that the physical processes at some point depend on the fields at other points (which gives rise to spatial dispersion) (e.g., [Serdyukov 01]):

$$\left\langle \vec{j} \right\rangle \left(\vec{r}, \omega\right) = -i\omega \int_V R_{LL}\left(\vec{r}, \vec{r}\,', \omega\right) \vec{E}\left(\vec{r}\,', \omega\right) dr' \tag{52}$$

This equation in case of translational invariance can be written as:

$$\left\langle \vec{j} \right\rangle \left(\vec{r}, \omega\right) = -i\omega \int_V R_{LL}\left(\vec{r} - \vec{r}\,', \omega\right) \vec{E}\left(\vec{r}\,', \omega\right) dr' \tag{53}$$

It has to be realized that (53) assumes translational symmetry of the considered object, which in turn means that (53) and any consequences cannot be directly applied to the consideration of the boundary condition problems. Transforming (53) to the spatial Fourier domain, we get:

$$\left\langle j \right\rangle_\alpha \left(\vec{k}, \omega\right) = -i\omega R_{LL,\alpha\beta}\left(\vec{k}, \omega\right) E_\beta\left(\vec{k}, \omega\right) \tag{54}$$

Form (54) is an expansion of the averaged current and fields over plane waves. In principle, the dependence of the averaged current on the magnetic field can be explicitly included in (54) as well. Nevertheless taking into account that in a plane wave the magnetic field can always be expressed through the electric field $B_\alpha = \frac{c}{\omega} e_{\alpha\beta l} k_\beta E_l$ ( $e_{\alpha\beta l}$ is the Levi-Chivita tensor), one can leave out the dependence on the magnetic field without loss of generality.

From (22) we obtain:

$$\begin{gathered} P_{LL,\alpha}\left(\vec{k}, \omega\right) = R_{LL,\alpha\beta}\left(\vec{k}, \omega\right) E_\beta\left(\vec{k}, \omega\right) \\ D_{LL,\alpha}\left(\vec{k}, \omega\right) = E_\alpha\left(\vec{k}, \omega\right) + 4\pi P_{LL,\alpha}\left(\vec{k}, \omega\right) = \left(\delta_{\alpha\beta} + 4\pi R_{LL,\alpha\beta}\left(\vec{k}, \omega\right)\right) E_\beta\left(\vec{k}, \omega\right) = \varepsilon_{LL,\alpha\beta}\left(\vec{k}, \omega\right) E_\beta\left(\vec{k}, \omega\right) \end{gathered} \tag{55}$$

Here the effective permittivity has been introduced:

$$\varepsilon_{LL,\alpha\beta}\left(\vec{k},\omega\right)=\delta_{\alpha\beta}+4\pi R_{LL,\alpha\beta}\left(\vec{k},\omega\right) \quad (56)$$

The respective dispersion relation is:

$$\det\left[e_{\alpha\beta l}k_\beta e_{l\beta\gamma}k_\beta+\frac{\omega^2}{c^2}\varepsilon_{LL,\alpha\beta}(\vec{k},\omega)\right]=0 \quad (57)$$

The introduced above function $\varepsilon_{LL,\alpha\beta}\left(\vec{k},\omega\right)$ - effective permittivity - is a tensor with components depending on both the frequency and the wave vector. It is worth noticing that the effective permittivity introduced this way depends not only on the metaatom properties but on the excitation conditions as well, und thus far it cannot be called "material parameter" – i.e., a parameter which depends on the material properties only (this issue will be discussed in details later).

Nevertheless, it should be stated that the knowledge of the effective permittivity fully solves the problem of propagation of plane waves in bulk media. Equation (57) can have several solutions for the same propagation direction and the same polarisation state [Pekar 82].

When we use this representation, it has to be clearly realized that:

1. We are working with the "L&L" representation where there is no magnetization (the magnetic response is included through spatial dispersion of the electric polarization (55)).
2. Form (52) assumes translational invariance of the media, which means that the form is acceptable for homogeneous material far from the boundaries (bulk materials).
3. In the framework of (52) it is impossible to introduce any permeability, because of in "L&L" representation there is no magnetization. All magnetic effects are included in the effective permittivity. In order to introduce permeability it is necessary to transit the "L&L" to the "C" form, which cannot be done unambiguously (see section 1.6 – "Transformation between different representations").

***Case of weak spatial dispersion***

The phenomenological model (52) is widely used in different branches of physics like plasma physics or physics of crystals. In the vast majority of the considered problems, the function $R\left(\vec{k},\omega\right)$ (or equivalently $\varepsilon_{LL}\left(\vec{k},\omega\right)$) is expanded into the Taylor series up to the second order, namely:

$$R_{LL,\alpha\beta}\left(\vec{k},\omega\right) \approx R_{LL,\alpha\beta}\left(\vec{k}_0,\omega\right) + \left.\frac{\partial R_{LL,\alpha\beta}\left(\vec{k},\omega\right)}{\partial k_\gamma}\right|_{\vec{k}=\vec{k}_0}\left(k_\gamma - k_{0,\gamma}\right) + \frac{1}{2}\left.\frac{\partial^2 R_{LL,\alpha\beta}\left(\vec{k},\omega\right)}{\partial k_\gamma \partial k_\delta}\right|_{\vec{k}=\vec{k}_0}\left(k_\gamma - k_{0,\gamma}\right)\left(k_\delta - k_{0,\delta}\right) \tag{58}$$

The functions $R_{LL,\alpha\beta}\left(\vec{k}_0,\omega\right), \left.\frac{\partial R_{LL,\alpha\beta}\left(\vec{k},\omega\right)}{\partial k_\gamma}\right|_{\vec{k}=\vec{k}_0}, \frac{1}{2}\left.\frac{\partial^2 R_{LL,\alpha\beta}\left(\vec{k},\omega\right)}{\partial k_\gamma \partial k_\delta}\right|_{\vec{k}=\vec{k}_0}$ are supposed to be found from experiments or rigorous microscopic calculations.

It should be pointed out that, expansion itself can be performed around any $\vec{k}_0$, not necessarily $\vec{k}_0 = 0$, provided that expansion over angles and wavelengths is properly done; in other words, the expansion formally can be written for small spatial dispersion and for strong spatial dispersion, but in the last case only for waves with the wave numbers close to $\vec{k}_0$. The math in this case does not impose any limitations.

After having all this said, the final form of the $R_{LL,\alpha\beta}$ in the "L&L" representation in case of weak spatial dispersion can be written as:

$$R_{LL,\alpha\beta}\left(\vec{k},\omega\right) \approx R_{LL,\alpha\beta}\left(0,\omega\right) + \left.\frac{\partial R_{LL,\alpha\beta}\left(\vec{k},\omega\right)}{\partial k_\gamma}\right|_{\vec{k}=0} k_\gamma + \frac{1}{2}\left.\frac{\partial^2 R_{LL,\alpha\beta}\left(\vec{k},\omega\right)}{\partial k_\gamma \partial k_\delta}\right|_{\vec{k}=0} k_\gamma k_\delta \tag{59}$$

and the respective relations are:

$$\begin{cases} P_{LL,\alpha}\left(\vec{k},\omega\right)=\left( R_{LL,\alpha\beta}\left(0,\omega\right)+\left.\dfrac{\partial R_{LL,\alpha\beta}\left(\vec{k},\omega\right)}{\partial k_\gamma}\right|_{\vec{k}=0} k_\gamma+\dfrac{1}{2}\left.\dfrac{\partial^2 R_{LL,\alpha\beta}\left(\vec{k},\omega\right)}{\partial k_\gamma \partial k_\delta}\right|_{\vec{k}=0} k_\gamma k_\delta \right) E_\beta\left(\vec{k},\omega\right) \\ D_{LL,\alpha}\left(\vec{k},\omega\right)=E_\alpha\left(\vec{k},\omega\right)+4\pi P_{LL,\alpha}\left(\vec{k},\omega\right)= \\ \qquad =\left(\varepsilon^{(0)}{}_{LL,\alpha\beta}\left(\omega\right)+\varepsilon^{(1)}{}_{LL,\alpha\beta\gamma}\left(\omega\right)k_\gamma+\varepsilon^{(2)}{}_{LL,\alpha\beta\gamma\delta}\left(\omega\right)k_\gamma k_\delta\right)E_\beta\left(\vec{k},\omega\right) \\ \varepsilon^{(0)}{}_{LL,\alpha\beta}\left(\omega\right)=\left(\delta_{\alpha\beta}+4\pi R_{LL,\alpha\beta}\left(0,\omega\right)\right) \\ \varepsilon^{(1)}{}_{LL,\alpha\beta\gamma}\left(\omega\right)=4\pi\left.\dfrac{\partial R_{LL,\alpha\beta}\left(\vec{k},\omega\right)}{\partial k_\gamma}\right|_{\vec{k}=0} \\ \varepsilon^{(2)}{}_{LL,\alpha\beta\gamma\delta}\left(\omega\right)=2\pi\left.\dfrac{\partial^2 R_{LL,\alpha\beta}\left(\vec{k},\omega\right)}{\partial k_\gamma \partial k_\delta}\right|_{\vec{k}=0} \\ \left\langle j\right\rangle_\alpha\left(\vec{k},\omega\right)=-\dfrac{i\omega}{4\pi}\left(\varepsilon^{(0)}{}_{LL,\alpha\beta}\left(\omega\right)-\delta_{\alpha\beta}+\varepsilon^{(1)}{}_{LL,\alpha\beta\gamma}\left(\omega\right)k_\gamma+\varepsilon^{(2)}{}_{LL,\alpha\beta\gamma\delta}\left(\omega\right)k_\gamma k_\delta\right)E_\beta\left(\vec{k},\omega\right) \end{cases} \tag{60}$$

We see that the expansion of function $R_{LL,\alpha\beta}\left(\vec{k},\omega\right)$ up to the second order corresponds to the expansion of the averaged current $\left\langle j\right\rangle_\alpha\left(\vec{k},\omega\right)$ up to the second order as well; the expression for the averaged current $\left\langle j\right\rangle_\alpha\left(\vec{k},\omega\right)$ is useful, because the averaged current is invariant in all representations (see (37) in case of untransformed fields, e.g. $\vec{T}_1=0$) and can be used to compare the weak dispersion expansions in all representations.

Now it is worth considering relation of expansion (57) to the other representations for the spatial dispersion, used in various publications.

Expansion (60) is rather well known [Agranovich 84], tensors $\varepsilon^{(0)}{}_{LL,\alpha\beta}\left(\omega\right)$, $\varepsilon^{(1)}{}_{LL,\alpha\beta\gamma}\left(\omega\right)$, and $\varepsilon^{(2)}{}_{LL,\alpha\beta\gamma\delta}\left(\omega\right)$ depend on the symmetry properties of the considered system and satisfy symmetry principles of Onsager coefficients.

It should be noted that in the framework of the “L&L” representation for the case of weak dispersion (up to the second order) three tensor functions - $\varepsilon^{(0)}{}_{LL,\alpha\beta}\left(\omega\right)$, $\varepsilon^{(1)}{}_{LL,\alpha\beta\gamma}\left(\omega\right)$, and $\varepsilon^{(2)}{}_{LL,\alpha\beta\gamma\delta}\left(\omega\right)$ - have been obtained, but the functions are not independent because they originate from the same function $R_{LL,\alpha\beta}\left(\vec{k},\omega\right)$.

### 2.1.2 “C” representation

***Case of strong spatial dispersion***

Basically instead of formal introduction of one function for polarizability (52), in case of “C” representation we have to introduce two functions for polarizability $P_{C,\alpha}$ and magnetization $M_{C,\alpha}$:

$$\begin{cases} P_{C,\alpha}\left(\vec{k},\omega\right)=R_{C,\alpha\beta}\left(\vec{k},\omega\right)E_{\beta}\left(\vec{k},\omega\right)+R'_{C,\alpha\beta}\left(\vec{k},\omega\right)B_{\beta}\left(\vec{k},\omega\right) \\ M_{C,\alpha}\left(\vec{k},\omega\right)=F'_{C,\alpha\beta}\left(\vec{k},\omega\right)E_{\beta}\left(\vec{k},\omega\right)+F_{C,\alpha\beta}\left(\vec{k},\omega\right)B_{\beta}\left(\vec{k},\omega\right) \end{cases} \tag{61}$$

which gives the most general form of expressions for $D_{C,\alpha}$:

$$D_{C,\alpha}\left(\vec{k},\omega\right)=\left(\delta_{\alpha\beta}+4\pi R_{C,\alpha\beta}\left(\vec{k},\omega\right)\right)E_{\beta}\left(\vec{k},\omega\right)+4\pi R'_{C,\alpha\beta}\left(\vec{k},\omega\right)B_{\beta}\left(\vec{k},\omega\right) \tag{62}$$

As for the magnetic material relations, the commonly assumed step would be an introduction of $H_{\alpha}=B_{\alpha}-4\pi M_{C,\alpha}$:

$$H_{C,\alpha}\left(\vec{k},\omega\right)=-4\pi F'_{C,\alpha\beta}\left(\vec{k},\omega\right)E_{\beta}\left(\vec{k},\omega\right)+\left(\delta_{\alpha\beta}+4\pi F_{C,\alpha\beta}\left(\vec{k},\omega\right)\right)B_{\beta}\left(\vec{k},\omega\right) \tag{63}$$

At this point it is worth noting that formally, using the respective transformations between the electric and magnetic fields $B_{\alpha}=\frac{c}{\omega}e_{\alpha\beta l}k_{\beta}E_{l}$, the expressions for $D_{C,\alpha}$ and $H_{C,\alpha}$ can be written as:

$$\begin{cases} D_{C,\alpha}\left(\vec{k},\omega\right)=\varepsilon_{C,\alpha\beta}\left(\vec{k},\omega\right)E_{\beta}\left(\vec{k},\omega\right) \\ P_{C,\alpha}\left(\vec{k},\omega\right)=\left(R_{C,\alpha\beta}\left(\vec{k},\omega\right)+\frac{c}{\omega}R'_{C,\alpha l}\left(\vec{k},\omega\right)e_{l\gamma\beta}k_{\gamma}\right)E_{\beta}\left(\vec{k},\omega\right) \\ \varepsilon_{C,\alpha\beta}\left(\vec{k},\omega\right)=\delta_{\alpha\beta}+4\pi R_{C,\alpha\beta}\left(\vec{k},\omega\right)+\frac{4\pi c}{\omega}R'_{C,\alpha l}\left(\vec{k},\omega\right)e_{l\gamma\beta}k_{\gamma} \\ M_{C,\alpha}\left(\vec{k},\omega\right)=\left(F'_{C,\alpha\beta}\left(\vec{k},\omega\right)+\frac{c}{\omega}F_{C,\alpha l}\left(\vec{k},\omega\right)e_{\alpha l\beta}k_{l}\right)E_{\beta} \\ H_{\alpha}\left(\vec{k},\omega\right)=\xi_{\alpha,\beta}\left(\vec{k},\omega\right)E_{\beta}\left(\vec{k},\omega\right) \\ \xi_{\alpha\beta}\left(\vec{k},\omega\right)=-\left(4\pi F'_{C,\alpha\beta}\left(\vec{k},\omega\right)+\frac{c}{\omega}\left(\delta_{\alpha l}+4\pi F_{C,\alpha l}\left(\vec{k},\omega\right)\right)\left(e_{l\gamma\beta}k_{\gamma}\right)\right) \end{cases} \tag{64a}$$

Using the representation (64a) it is rather straightforward to write down a "propagation equation" – analogy of the well-established Helmholz equation for the plane electromagnetic wave propagation [Jackson 99]. The last equation in (26) after substitution $D_{C,\alpha}\left(\vec{k},\omega\right)$ and $H_{C,\alpha}\left(\vec{k},\omega\right)$ from (64a) becomes a "propagation equation":

$$e_{\alpha\beta\gamma}k_{\beta}H_{\gamma}=D_{\alpha}\Rightarrow e_{\alpha\beta\gamma}k_{\beta}\xi_{\gamma p}E_{p}=\varepsilon_{\alpha\beta}E_{\beta}\Rightarrow \det\left|e_{\alpha p\gamma}k_{p}\xi_{\gamma\beta}-\varepsilon_{\alpha\beta}\right|=0 \tag{64b}$$

Propagation equation and respective dispersion relation (64b) appear to be much more simple and natural in compare with the usually used Helmholz equation; moreover form (64a) does not allow us to introduce a kind of permeability $\mu$ straightforwardly. In order to introduce $\mu$, instead of substitution the magnetic field $\vec{B}$ in terms of the electric field $\vec{E}$, we should perform the opposite operation and express $\vec{E}$ through $\vec{B}$. In order to do it, we have to solve the equation

$$\left[\vec{k}\times\vec{E}\right]=-\frac{\omega}{c}\vec{B} \tag{65}$$

assuming that the wave vector $\vec{k}$ and the magnetic field $\vec{B}$ are known and considering the electric field $\vec{E}$ as a variable. From the vector analysis solution of this problem is known, namely if there are three vectors $\vec{x}$, $\vec{a}$, and $\vec{b}$ so that $\left[\vec{x}\times\vec{a}\times\vec{b}\right]\neq 0$ and $\left[\vec{x}\times\vec{a}\right]=\vec{b}$ then the solution for $\vec{x}$ is:

$$\begin{cases}\vec{x}=\vec{a}\dfrac{\gamma}{|a|^{2}}+\left[\vec{a}\times\vec{b}\right]\dfrac{1}{|a|^{2}}\\ \gamma=\left(\vec{x}\times\vec{a}\right)\end{cases} \tag{66}$$

or, in other words, this requires the knowledge of one more constant $\gamma$ .

In our case the last requirement is given by the Maxwell equation $\left(\vec{k}\times\vec{E}\right)=-4\pi\left(\vec{k}\times\vec{P}\right)$ and the final solution of (65) is:

$$\vec{E}=-\vec{k}\frac{4\pi\left(\vec{k}\times\vec{P}\right)}{|k|^{2}}+\frac{\omega}{c}\left[\vec{k}\times\vec{B}\right]\frac{1}{|k|^{2}} \tag{67}$$

or, in terms of vector components:

$$E_\alpha = -4\pi k_\alpha \left( \frac{k_i P_{C,i}}{k_j k_j^*} \right) + \frac{\omega}{c} e_{\alpha\beta\gamma} k_\beta B_\gamma \left( \frac{1}{k_m k_m^*} \right) \tag{68}$$

Substituting (68) into (63) we have:

$$\begin{aligned} H_{C,\alpha}\left(\vec{k},\omega\right) &= -\left(4\pi\right)^2 F'_{C,\alpha\beta}\left(\vec{k},\omega\right) k_\beta \left( \frac{k_i P_{C,i}}{k_j k_j} \right) + \\ &+ \left[ \left( \delta_{\alpha\beta} + 4\pi F_{C,\alpha\beta}\left(\vec{k},\omega\right) \right) + \frac{4\pi\omega}{c} F'_{C,\alpha\varsigma}\left(\vec{k},\omega\right) e_{\varsigma\gamma\beta} k_\gamma \left( \frac{1}{k_m k_m^*} \right) \right] B_\beta\left(\vec{k},\omega\right) \end{aligned} \tag{69}$$

It is clear that in this case any attempt to introduce proportionality between $H_{C,\alpha}$ and $B_\alpha$ in form $B_\alpha = \mu_{\alpha\beta}\left(\vec{k},\omega\right) H_{C,\alpha}$ fails if polarisation $P_{C,\alpha}$ and the wave vector $k_\alpha$ are not perpendicular to each other $k_\alpha P_{C,\alpha} \neq 0$. In general, the polarizability $P_{C,\alpha}$ is not perpendicular to $k_\alpha$ and permeability cannot be introduced at all. It is worth noting that the problem arises from the fact that the magnetic response is stipulated by an interaction with the electric field, not with the magnetic one. Obviously, introduction of magnetic constant in form of a proportionality coefficient between $H_{C,\alpha}$ and $B_\alpha$ is neither logical nor necessary – the form (64a) is much more physically justified than the form $B_\alpha = \mu_{\alpha\beta}\left(\vec{k},\omega\right) H_{C,\alpha}$, which is unconditionally suitable only for the case of interaction of a system with the magnetic field. Nevertheless, in case $k_\alpha P_{C,\alpha} = 0$ the electric field can be unambiguously and straightforwardly presented as a function of the magnetic field $E_\alpha = \frac{\omega}{c} e_{\alpha\beta\gamma} k_\beta B_\gamma \left( \frac{1}{k_m k_m^*} \right)$ and (69) can be rewritten as:

$$\begin{cases} H_{C,\alpha}\left(\vec{k},\omega\right) = \left[ \mu_{\alpha\beta}\left(\vec{k},\omega\right) \right]^{-1} B_\beta\left(\vec{k},\omega\right) \\ \left[ \mu_{\alpha\beta}\left(\vec{k},\omega\right) \right]^{-1} = \left( \delta_{\alpha\beta} + 4\pi F_{C,\alpha\beta}\left(\vec{k},\omega\right) \right) + \frac{4\pi\omega}{c} F'_{C,\alpha\varsigma}\left(\vec{k},\omega\right) e_{\varsigma\gamma\beta} k_\gamma \left( \frac{1}{k_m k_m^*} \right) \end{cases} \tag{70}$$

It has to be noted, that even in this case (when the proportionality between $H_{C,\alpha}$ and $B_\alpha$ can be established), introduction of $\left[\mu_{\alpha\beta}\left(\vec{k},\omega\right)\right]^{-1}$ (not $\mu_{\alpha\beta}\left(\vec{k},\omega\right)$!) appears to be the logical step in the elaboration of the homogenization model. Here it is seen also, that if magnetization is caused by the electric field ($F'_{C,\alpha\varsigma}\left(\vec{k},\omega\right)\neq 0$), then the introduced this way $\mu_{\alpha\beta}\left(\vec{k},\omega\right)$ is spatially dispersive even in case of non-spatially dispersive response $F'_{C,\alpha\beta}\left(\vec{k},\omega\right)=F'_{C,\alpha\beta}\left(\omega\right)$. In fact, as it will be clear later from the consideration of the multipole model $F'_{C}\left(\vec{k},\omega\right)\sim k$ , which results in absence of spatial dispersion for $\mu_{\alpha\beta}\left(\vec{k},\omega\right)=\mu_{\alpha\beta}\left(\omega\right)$ under weak (up to the second order) spatial dispersion approximation.

Considering a plane wave propagating into $z$ direction in a media with tensor character of polarizability, one can easily see that in order to satisfy the condition $k_\alpha P_{C,\alpha}=0$ (which guarantees a possibility to introduce the permeability) we have to have tensor $\varepsilon_{C,\alpha\beta}\left(\vec{k},\omega\right)$ in the following form:

$$\begin{pmatrix} \varepsilon_{C,xx}\left(\vec{k},\omega\right) & \varepsilon_{C,xy}\left(\vec{k},\omega\right) & 0 \\ \varepsilon_{C,yx}\left(\vec{k},\omega\right) & \varepsilon_{C,yy}\left(\vec{k},\omega\right) & 0 \\ 0 & 0 & 0 \end{pmatrix} \qquad (71)$$

It is interesting to demonstrate a design which, according to (71) does not allow introducing of the permeability due to the appearance of the polarization $P_{C,\alpha}$, parallel to the wave vector $k_\alpha$, so that $k_\alpha P_{C,\alpha}\neq 0$. This could be a SRR structure placed with its top part parallel to the wave vector $k_\alpha$, as it is shown in Fig. 5.

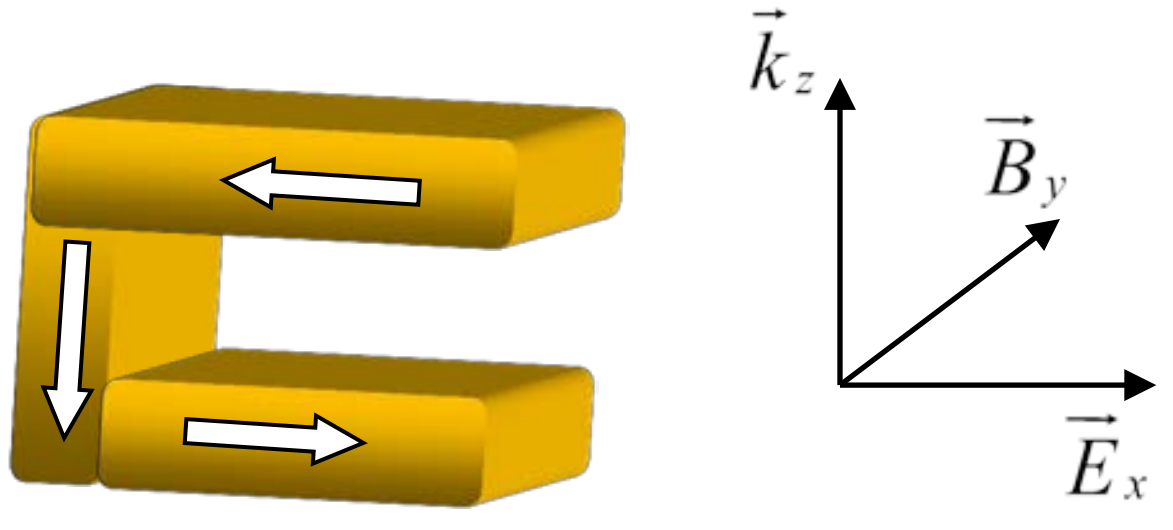

**Fig. 5: Design and positioning of a metaatom which does not allow introducing of permeability. Appearance of polarization parallel to the wave vector is caused by the short cut between parts of the metaatom.**

Now it is methodologically interesting to find relations between “C” and “L&L” representation in the frame of the strong spatial dispersion. It can be easily performed by equating the relations for the averaged current in both representations (let us remind, that according to SFT the averaged current is not changed provided fields remain unchanged as well), which gives after some algebra:

$$R_{LL,\alpha\beta}\left(\vec{k},\omega\right)=R_{C,\alpha\beta}\left(\vec{k},\omega\right)+\frac{c}{\omega}\left(R'_{C,\alpha\beta}\left(\vec{k},\omega\right)e_{l\gamma\beta}+F'_{C,l\beta}\left(\vec{k},\omega\right)e_{\alpha\gamma l}\right)k_{\gamma}+\left(\frac{c}{\omega}\right)^{2}F_{C,\gamma l}\left(\vec{k},\omega\right)e_{\alpha m\gamma}e_{lp\beta}k_{m}k_{\beta} \tag{72}$$

It is easy to see that even in the case of dispersion-free response in “C” representation ( $R_{C,\alpha\beta}\left(\vec{k},\omega\right)=R_{C,\alpha\beta}\left(\omega\right)$, $R'_{C,\alpha\beta}\left(\vec{k},\omega\right)=R'_{C,\alpha\beta}\left(\omega\right)$, $F'_{C,l\beta}\left(\vec{k},\omega\right)=F'_{C,l\beta}\left(\omega\right)$, $F_{C,\gamma l}\left(\vec{k},\omega\right)=F_{C,\gamma l}\left(\omega\right)$ ), spatial dispersion unavoidably appears in “L&L” representation. Moreover, if “non-eigen” responses are not zero ( $R'_{C,\alpha\beta}\left(\vec{k},\omega\right)\neq 0$, $F'_{C,l\beta}\left(\vec{k},\omega\right)\neq 0$ ), then $R_{LL,\alpha\beta}\left(\vec{k},\omega\right)$ is proportional at least to the first order of the wave vector (linear spatial dispersion); if the magnetization is caused by an interaction with the magnetic field ( $F_{C,l\beta}\left(\vec{k},\omega\right)\neq 0$ ) then $R_{LL,\alpha\beta}\left(\vec{k},\omega\right)$ is proportional at least to the second order of the wave vector (quadratic spatial dispersion). Thus, one can conclude that the nature of interaction (dependence of the polarizability and magnetization on the electric and/or magnetic fields) is the basic question, which determines the whole theoretical construction of the homogenization of the Maxwell equations. The nature of interaction of the electromagnetic wave with the charges in metallic nanoresonator (in case of MMs based on plasmonic structures, for example in the optical domain) is the interaction of electrons with the electric field only, because typical velocities of electrons are far from the velocity of light:

$$\frac{d\vec{p}_i}{dt}=q_i\vec{e}+\frac{q_i}{c}\left[\vec{v}_i*\vec{h}\right]\approx q_i\vec{e} \tag{73}$$

Here $\vec{e}$ and $\vec{h}$ are the microscopic electric and magnetic fields, respectively, $\vec{q_i}$, $\vec{p_i}$ and $\vec{v_i}$ are the charges, pulses, and velocities of charges, $c$ is the speed of light. Hence there is no reason to assume that the magnetic field should appear in the phenomenological expression for the polarizability and magnetization, i.e. $R'_{C,\alpha\beta}\left(\vec{k},\omega\right)=0$, $F_{C,\gamma l}\left(\vec{k},\omega\right)=0$. In this case the elaborated above expressions can be summarized one more time:

$$\begin{cases} P_{C,\alpha}\left(\vec{k},\omega\right)=R_{C,\alpha\beta}\left(\vec{k},\omega\right)E_{\beta}\left(\vec{k},\omega\right) \\ M_{C,\alpha}\left(\vec{k},\omega\right)=F'_{C,\alpha\beta}\left(\vec{k},\omega\right)E_{\beta}\left(\vec{k},\omega\right) \end{cases} \tag{74}$$

$$\begin{cases} D_{C,\alpha}\left(\vec{k},\omega\right)=\varepsilon_{C,\alpha\beta}\left(\vec{k},\omega\right)E_{\beta}\left(\vec{k},\omega\right) \\ \varepsilon_{C,\alpha\beta}\left(\vec{k},\omega\right)=\delta_{\alpha\beta}+4\pi R_{C,\alpha\beta}\left(\vec{k},\omega\right) \\ H_{\alpha}\left(\vec{k},\omega\right)=\xi_{\alpha\beta}\left(\vec{k},\omega\right)E_{\beta}\left(\vec{k},\omega\right) \\ \xi_{\alpha\beta}\left(\vec{k},\omega\right)=-\left(4\pi F'_{C,\alpha\beta}\left(\vec{k},\omega\right)+\dfrac{c}{\omega}e_{\alpha\gamma\beta}k_{\gamma}\right) \end{cases} \tag{75}$$

$$\begin{cases} D_{C,\alpha}\left(\vec{k},\omega\right)=\varepsilon_{C,\alpha\beta}\left(\vec{k},\omega\right)E_{\beta}\left(\vec{k},\omega\right) \\ \varepsilon_{C,\alpha\beta}\left(\vec{k},\omega\right)=\delta_{\alpha\beta}+4\pi R_{C,\alpha\beta}\left(\vec{k},\omega\right) \\ H_{C,\alpha}\left(\vec{k},\omega\right)==\left[\mu_{\alpha\beta}\left(\vec{k},\omega\right)\right]^{-1}B_{\beta}\left(\vec{k},\omega\right) \\ \left[\mu_{\alpha\beta}\left(\vec{k},\omega\right)\right]^{-1}=\left[\delta_{\alpha\beta}+\dfrac{4\pi\omega}{c}F'_{C,\alpha\varsigma}\left(\vec{k},\omega\right)e_{\varsigma\gamma\beta}k_{\gamma}\left(\dfrac{1}{k_m k_m^*}\right)\right] \end{cases} \tag{76}$$

In the scientific literature there is a widely known and used representation of permittivity in the following form ([Landau&Lifshits], [Vinogradov 01]):

$$\varepsilon_{C,\alpha\beta}(\vec{k},\omega)=\varepsilon_C^{tr}(\vec{k},\omega)\left(\delta_{\alpha\beta}-\frac{k_\alpha k_\beta}{k^2}\right)+\varepsilon_C^{l}(\vec{k},\omega)\frac{k_\alpha k_\beta}{k^2} \tag{77}$$

which is basically a special case of (75) or (76).
There is a hypothesis [Turov 83] that the permeability obeys the same type of expression:

$$\mu_{C,\alpha\beta}(\vec{k},\omega)=\mu_C^{tr}(\vec{k},\omega)\left(\delta_{\alpha\beta}-\frac{k_\alpha k_\beta}{k^2}\right)+\mu_C^{l}(\vec{k},\omega)\frac{k_\alpha k_\beta}{k^2} \tag{78}$$

Exhausting consideration of relations between the coefficients $\varepsilon_C^{tr}(\omega)$, $\varepsilon_C^{l}(\omega)$, $\mu_C^{tr}(\omega)$, $\mu_C^{l}(\omega)$ from one side and $\varepsilon_{LL}^{tr}(\omega)$, and $\varepsilon_{LL}^{tr}(\omega)$ from another side can be found in [Vinogradov 01].

It is necessary to mention, that some authors try to subdivide function $P_{C,\alpha}\left(\vec{k},\omega\right)$ on a dipole and a quadrupole parts within the frame of the phenomenologic approach. This option is supported by a proven in [Vinogradov 99] possibility to express the averaged current in the "C" representation through the dipole, quadrupole, and magnetic dipole parts, namely:

$$\left\langle j\right\rangle_{\alpha}=-i\omega\left(P_{C,\alpha}-k_{\beta}Q_{\alpha\beta}\right)+ice_{\alpha\beta l}k_{\beta}M_{C,l} \quad (79)$$

Here $P_{C,\alpha}$, $Q_{\alpha\beta}$, and $M_{C,l}$ are the dipole, qudrupole, and magnetic dipole contributions (see (14), (15)). After that the three mentioned parts have to be reformulated in analogy with (61) through electric and (possibly) magnetic fields, which leads us finally to the same kind of expressions for $D_{C,\alpha}$ and $H_{\alpha}$ as in (75) and (76). It is seen, that in the framework of the phenomenological approach it does not make sense to subdivide function $P_{C,\alpha}\left(\vec{k},\omega\right)$ into dipole and quadrupole parts. In order to benefit from the representation (79) it is necessary to have a functional form for all three parts $P_{C,\alpha}$, $Q_{\alpha\beta}$, and $M_{C,l}$ [Bosi 82], [Graham 96], [Raab 94], which is not possible within the frameworks of the phenomenological model; an attempt to get these functions will be performed in the next chapter 2.2.

***Case of weak spatial dispersion***

In the case of weak spatial dispersion functions $R_{C,\alpha\beta}$, $R'_{C,\alpha\beta}$, $F_{C,\alpha\beta}$, $F'_{C,\alpha\beta}$ can be expanded into the Taylor series (we limit our consideration by the second order of expansion):

$$\begin{cases} P_{C,\alpha}\left(\vec{k},\omega\right)=R_{C,\alpha\beta}\left(\vec{k},\omega\right)E_{\beta}\left(\vec{k},\omega\right)+R'_{C,\alpha\beta}\left(\vec{k},\omega\right)B_{\beta}\left(\vec{k},\omega\right) \\ M_{C,\alpha}\left(\vec{k},\omega\right)=F'_{C,\alpha\beta}\left(\vec{k},\omega\right)E_{\beta}\left(\vec{k},\omega\right)+F_{C,\alpha\beta}\left(\vec{k},\omega\right)B_{\beta}\left(\vec{k},\omega\right) \\ R_{C,\alpha\beta}\left(\vec{k},\omega\right)\approx R_{C,\alpha\beta}\left(0,\omega\right)+\left.\frac{\partial R_{C,\alpha\beta}\left(\vec{k},\omega\right)}{\partial k_{\gamma}}\right|_{\vec{k}=0}k_{\gamma}+\frac{1}{2}\left.\frac{\partial^{2}R_{C,\alpha\beta}\left(\vec{k},\omega\right)}{\partial k_{\gamma}\partial k_{\delta}}\right|_{\vec{k}=0}k_{\gamma}k_{\delta} \\ R'_{C,\alpha\beta}\left(\vec{k},\omega\right)\approx R'_{C,\alpha\beta}\left(0,\omega\right)+\left.\frac{\partial R'_{C,\alpha\beta}\left(\vec{k},\omega\right)}{\partial k_{\gamma}}\right|_{\vec{k}=0}k_{\gamma}+\frac{1}{2}\left.\frac{\partial^{2}R'_{C,\alpha\beta}\left(\vec{k},\omega\right)}{\partial k_{\gamma}\partial k_{\delta}}\right|_{\vec{k}=0}k_{\gamma}k_{\delta} \\ F_{C,\alpha\beta}\left(\vec{k},\omega\right)\approx F_{C,\alpha\beta}\left(0,\omega\right)+\left.\frac{\partial F_{C,\alpha\beta}\left(\vec{k},\omega\right)}{\partial k_{\gamma}}\right|_{\vec{k}=0}k_{\gamma}+\frac{1}{2}\left.\frac{\partial^{2}F_{C,\alpha\beta}\left(\vec{k},\omega\right)}{\partial k_{\gamma}\partial k_{\delta}}\right|_{\vec{k}=0}k_{\gamma}k_{\delta} \\ F'_{C,\alpha\beta}\left(\vec{k},\omega\right)\approx F'_{C,\alpha\beta}\left(0,\omega\right)+\left.\frac{\partial F'_{C,\alpha\beta}\left(\vec{k},\omega\right)}{\partial k_{\gamma}}\right|_{\vec{k}=0}k_{\gamma}+\frac{1}{2}\left.\frac{\partial^{2}F'_{C,\alpha\beta}\left(\vec{k},\omega\right)}{\partial k_{\gamma}\partial k_{\delta}}\right|_{\vec{k}=0}k_{\gamma}k_{\delta} \end{cases} \tag{80}$$

which results for the respective expressions for $D_{C,\alpha}$ and $H_{\alpha}$. It is also appropriate at this point to mention that it is enough to restrict the consideration by the first order in the expansion for $F_{C,\alpha\beta}$ and $F'_{C,\alpha\beta}$, because after substitution to MEs the first order of spatial dispersion corresponds to the second order of the other part due to the fact, that the expansion is placed under curl operator, that in turn brings one more order of spatial dispersion:

$$\begin{cases} D_{C,\alpha}\left(\vec{k},\omega\right)=\left(\varepsilon^{(0)}_{C,\alpha\beta}\left(\omega\right)+\varepsilon^{(1)}_{C,\alpha\beta\gamma}\left(\omega\right)k_{\gamma}+\varepsilon^{(2)}_{C,\alpha\beta\gamma\delta}\left(\omega\right)k_{\gamma}k_{\delta}\right)E_{\beta}\left(\vec{k},\omega\right)+ \\ \qquad +\left(\psi^{(0)}_{C,\alpha\beta}\left(\omega\right)+\psi^{(1)}_{C,\alpha\beta\gamma}\left(\omega\right)k_{\gamma}+\psi^{(2)}_{C,\alpha\beta\gamma\delta}\left(\omega\right)k_{\gamma}k_{\delta}\right)B_{\beta}\left(\vec{k},\omega\right) \\ \\ \varepsilon^{(0)}_{C,\alpha\beta}\left(\omega\right)=\delta_{\alpha\beta}+4\pi R_{C,\alpha\beta}\left(0,\omega\right); \qquad \psi^{(0)}_{C,\alpha\beta}\left(\omega\right)=\delta_{\alpha\beta}+4\pi R'_{C,\alpha\beta}\left(0,\omega\right) \\ \varepsilon^{(1)}_{C,\alpha\beta\gamma}\left(\omega\right)=4\pi\left.\frac{\partial R_{C,\alpha\beta}\left(\vec{k},\omega\right)}{\partial k_{\gamma}}\right|_{\vec{k}=0}; \qquad \psi^{(1)}_{C,\alpha\beta\gamma}\left(\omega\right)=4\pi\left.\frac{\partial R'_{C,\alpha\beta}\left(\vec{k},\omega\right)}{\partial k_{\gamma}}\right|_{\vec{k}=0} \\ \varepsilon^{(2)}_{C,\alpha\beta\gamma\delta}\left(\omega\right)=2\pi\left.\frac{\partial^{2}R_{C,\alpha\beta}\left(\vec{k},\omega\right)}{\partial k_{\gamma}\partial k_{\delta}}\right|_{\vec{k}=0}; \qquad \psi^{(2)}_{C,\alpha\beta\gamma\delta}\left(\omega\right)=2\pi\left.\frac{\partial^{2}R_{C,\alpha\beta}\left(\vec{k},\omega\right)}{\partial k_{\gamma}\partial k_{\delta}}\right|_{\vec{k}=0} \end{cases} \tag{81}$$

$$\begin{cases} H_\alpha\left(\vec{k},\omega\right)=\left(\phi^{(0)}_{C,\alpha\beta}\left(\omega\right)+\phi^{(1)}_{C,\alpha\beta\gamma}\left(\omega\right)k_{\gamma\ \delta}\right)E_\beta\left(\vec{k},\omega\right)+\left(\mu^{(0)}_{C,\alpha\beta}\left(\omega\right)+\mu^{(1)}_{C,\alpha\beta\gamma}\left(\omega\right)k_\gamma\right)B_\beta\left(\vec{k},\omega\right) \\ \phi^{(0)}_{C,\alpha\beta}\left(\omega\right)=-4\pi F'_{C,\alpha\beta}\left(0,\omega\right); \qquad \mu^{(0)}_{C,\alpha\beta}\left(\omega\right)=\delta_{\alpha\beta}+4\pi F_{C,\alpha\beta}\left(0,\omega\right) \\ \phi^{(1)}_{C,\alpha\beta\gamma}\left(\omega\right)=-4\pi\left.\frac{\partial F'_{C,\alpha\beta}\left(\vec{k},\omega\right)}{\partial k_\gamma}\right|_{\vec{k}=0}; \qquad \mu^{(1)}_{C,\alpha\beta\gamma}\left(\omega\right)=4\pi\left.\frac{\partial F_{C,\alpha\beta}\left(\vec{k},\omega\right)}{\partial k_\gamma}\right|_{\vec{k}=0} \end{cases} \tag{82}$$

Depending on a particular situation one or more terms in this expansion can be set to zero, and relations between some terms can be established based on the symmetry conditions. Thus far, in order to complete the homogenization model in case of weak spatial dispersion it is necessary to fill in the following 3x4 matrix:

$$\begin{pmatrix} R_{C,\alpha\beta}\left(0,\omega\right) & \left.\frac{\partial R_{C,\alpha\beta}\left(\vec{k},\omega\right)}{\partial k_\gamma}\right|_{\vec{k}=0} & \left.\frac{\partial^2 R_{C,\alpha\beta}\left(\vec{k},\omega\right)}{\partial k_\gamma \partial k_\delta}\right|_{\vec{k}=0} \\ R'_{C,\alpha\beta}\left(0,\omega\right) & \left.\frac{\partial R'_{C,\alpha\beta}\left(\vec{k},\omega\right)}{\partial k_\gamma}\right|_{\vec{k}=0} & \left.\frac{\partial^2 R'_{C,\alpha\beta}\left(\vec{k},\omega\right)}{\partial k_\gamma \partial k_\delta}\right|_{\vec{k}=0} \\ F_{C,\alpha\beta}\left(0,\omega\right) & \left.\frac{\partial F_{C,\alpha\beta}\left(\vec{k},\omega\right)}{\partial k_\gamma}\right|_{\vec{k}=0} & 0 \\ F'_{C,\alpha\beta}\left(0,\omega\right) & \left.\frac{\partial F'_{C,\alpha\beta}\left(\vec{k},\omega\right)}{\partial k_\gamma}\right|_{\vec{k}=0} & 0 \end{pmatrix} \tag{83}$$

or, in terms of other notations:

$$\begin{pmatrix} \varepsilon^{(0)}_{C,\alpha\beta}\left(\omega\right) & \varepsilon^{(1)}_{C,\alpha\beta\gamma}\left(\omega\right) & \varepsilon^{(2)}_{C,\alpha\beta\gamma\delta}\left(\omega\right) \\ \psi^{(0)}_{C,\alpha\beta}\left(\omega\right) & \psi^{(1)}_{C,\alpha\beta\gamma}\left(\omega\right) & \psi^{(2)}_{C,\alpha\beta\gamma\delta}\left(\omega\right) \\ \mu^{(0)}_{C,\alpha\beta}\left(\omega\right) & \mu^{(1)}_{C,\alpha\beta\gamma}\left(\omega\right) & 0 \\ \phi^{(0)}_{C,\alpha\beta}\left(\omega\right) & \phi^{(1)}_{C,\alpha\beta\gamma}\left(\omega\right) & 0 \end{pmatrix} \tag{84}$$

For example, in [Simovski 10] the authors arrive to the following representation based on qualitative consideration of physical processes appearing at the interaction of the electromagnetic field with metaatoms:

$$\begin{pmatrix} \varepsilon^{(0)}_{C,\alpha\beta}(\omega) & 0 & 0 \\ \psi^{(0)}_{C,\alpha\beta}(\omega) & 0 & 0 \\ \mu^{(0)}_{C,\alpha\beta}(\omega) & 0 & 0 \\ \phi^{(0)}_{C,\alpha\beta}(\omega) & \phi^{(1)}_{C,\alpha\beta\gamma}(\omega) & 0 \end{pmatrix} \qquad (85a)$$

The coupling effect, described by tensors $\psi^{(0)}{}_{\alpha\beta} = -\phi^{(0)}{}_{\alpha\beta}$ is known in electromagnetism and is called bi-anisotropy [Serdyukov 01]; the tensor is called magnetoelectric coupling parameter [Simovski 03]. Two special cases of bianisotropic media are known: chirial media when tensor $\psi^{(0)}{}_{\alpha\beta}$ is symmetric and omega media when tensor $\psi^{(0)}{}_{\alpha\beta}$ is antisymmetric. More details about classification of different media based on introduced above model can be found in [Tretyakov 98]. The form of material equations accepted in [Simovski 10] is similar to the one presented in [Kriegler 10], where matrix (84) is written in the following form:

$$\begin{pmatrix} \varepsilon^{(0)}_{C,\alpha\beta}(\omega) & 0 & 0 \\ \psi^{(0)}_{C,\alpha\beta}(\omega) & 0 & 0 \\ \mu^{(0)}_{C,\alpha\beta}(\omega) & 0 & 0 \\ \phi^{(0)}_{C,\alpha\beta}(\omega) & 0 & 0 \end{pmatrix} \qquad (85b)$$

and differs from (85a) by the absence of linear spatial dispersion in the equation for magnetization, which is equivalent to neglecting of anisotropy.

In both papers [Simovski 10] and [Kriegler 10] and many others the functional forms of the expressions for $P_\alpha$ and $M_\alpha$ are introduced based not on the developed here phenomenological approach, but using the multipole approach, considered in the next chapter. The usual way is to calculate the dipole and magnetic dipole moments using expressions known from electrostatics. It is believed, that in applications in the optical domain this approach has at least three drawbacks, namely:

1. It is clear from the physical point of view, that magnetic field does not affect charge dynamics and has not to be included in the basic considerations; it has been shown above, that the introduction of response to magnetic field through the electric one using one of the Maxwell equations is not straightforward.

2. Authors usually do not distinguish local and averaged fields when the charge dynamics in metaatoms is considered, that hides in some cases the role of spatial dispersion.
3. Practically all authors, taking into account magnetic moment, do not include into consideration quadrupole moment, which is in most cases mandatory due to the fact that magnetic moment and quadrupole moment are of the same order of the multipole expansion. Negligence of the quadrupole moment leads in turn to incorrect material equation representation, for example artificial exclusion of the first-order spatial dispersion term. This in turn excludes from the consideration some effects of anisotropy. The quadrupole moment effects can be neglected as compared with the magnetic moment influence only for specific geometries of inclusions. For example, the fundamental mode of double split ring resonators widely used in microwave metamaterials is characterized by a strong magnetic moment but negligible electric quadrupole moment, because the total current along two rings is nearly uniform around the whole ring structure.

Below we try to present a more consistent way of introduction of material equations in the frame of the phenomenological approach. As it was mentioned above, magnetization for the considered MM is evidently proportional only to the electric field and its spatial derivatives – see (74)-(76). In this case of weak spatial dispersion the summarized expressions are:

$$\begin{cases} D_{C,\alpha}\left(\vec{k},\omega\right)=\left(\varepsilon^{(0)}_{C,\alpha\beta}\left(\omega\right)+\varepsilon^{(1)}_{C,\alpha\beta\gamma}\left(\omega\right)k_\gamma+\varepsilon^{(2)}_{C,\alpha\beta\gamma\delta}\left(\omega\right)k_\gamma k_\delta\right)E_\beta\left(\vec{k},\omega\right) \\ \varepsilon^{(0)}_{C,\alpha\beta}\left(\omega\right)=\delta_{\alpha\beta}+4\pi R_{C,\alpha\beta}\left(0,\omega\right); \\ \varepsilon^{(1)}_{C,\alpha\beta\gamma}\left(\omega\right)=4\pi\left.\dfrac{\partial R_{C,\alpha\beta}\left(\vec{k},\omega\right)}{\partial k_\gamma}\right|_{\vec{k}=0}; \\ \varepsilon^{(2)}_{C,\alpha\beta\gamma\delta}\left(\omega\right)=2\pi\left.\dfrac{\partial^2 R_{C,\alpha\beta}\left(\vec{k},\omega\right)}{\partial k_\gamma\partial k_\delta}\right|_{\vec{k}=0}; \end{cases} \tag{86}$$

$$\begin{cases} H_\alpha\left(\vec{k},\omega\right)=\left(\phi^{(0)}_{C,\alpha\beta}\left(\omega\right)+\phi^{(1)}_{C,\alpha\beta\gamma}\left(\omega\right)k_\gamma\right)E_\beta\left(\vec{k},\omega\right) \\ \phi^{(0)}_{C,\alpha\beta}\left(\omega\right)=\delta_{\alpha\beta}+4\pi F'_{C,\alpha\beta}\left(0,\omega\right) \\ \phi^{(1)}_{C,\alpha\beta\gamma}\left(\omega\right)=4\pi\left.\dfrac{\partial F'_{C,\alpha\beta}\left(\vec{k},\omega\right)}{\partial k_\gamma}\right|_{\vec{k}=0} \end{cases} \tag{87}$$

Hence, in order to complete the homogenization model it is necessary to fill in the following 3x2 matrix:

$$\begin{pmatrix} R_{C,\alpha\beta}\left(0,\omega\right) & \left.\dfrac{\partial R_{C,\alpha\beta}\left(\vec{k},\omega\right)}{\partial k_\gamma}\right|_{\vec{k}=0} & \left.\dfrac{\partial^2 R_{C,\alpha\beta}\left(\vec{k},\omega\right)}{\partial k_\gamma \partial k_\delta}\right|_{\vec{k}=0} \\ F'_{C,\alpha\beta}\left(0,\omega\right) & \left.\dfrac{\partial F'_{C,\alpha\beta}\left(\vec{k},\omega\right)}{\partial k_\gamma}\right|_{\vec{k}=0} & 0 \end{pmatrix} \tag{88}$$

or, in terms of other notations:

$$\begin{pmatrix} \varepsilon^{(0)}_{C,\alpha\beta}\left(\omega\right) & \varepsilon^{(1)}_{C,\alpha\beta\gamma}\left(\omega\right) & \varepsilon^{(2)}_{C,\alpha\beta\gamma\delta}\left(\omega\right) \\ \phi^{(0)}_{C,\alpha\beta}\left(\omega\right) & \phi^{(1)}_{C,\alpha\beta\gamma}\left(\omega\right) & 0 \end{pmatrix} \tag{89}$$

Referring again to the paper [Simovski 10] one can conclude that the suggested there representation is equivalent to:

$$\begin{pmatrix} \varepsilon^{(0)}_{C,\alpha\beta}\left(\omega\right) & 0 & \varepsilon^{(2)}_{C,\alpha\beta\gamma\delta}\left(\omega\right) \\ \phi^{(0)}_{C,\alpha\beta}\left(\omega\right) & \phi^{(1)}_{C,\alpha\beta\gamma}\left(\omega\right) & 0 \end{pmatrix} \tag{90}$$

and the representation accepted in [Kriegler 10] is equivalent to:

$$\begin{pmatrix} \varepsilon^{(0)}_{C,\alpha\beta}\left(\omega\right) & 0 & \varepsilon^{(2)}_{C,\alpha\beta\gamma\delta}\left(\omega\right) \\ \phi^{(0)}_{C,\alpha\beta}\left(\omega\right) & 0 & 0 \end{pmatrix} \tag{91}$$

### 2.1.3 Transformation between "C" and "L&L" representations in case of strong spatial dispersion

The relation between "L&L" and "C" representations can be obtained from (72) taking into account the fact that the direct interaction with the magnetic field is absent:

$$R_{LL,\alpha\beta}\left(\vec{k},\omega\right)=R_{C,\alpha\beta}\left(\vec{k},\omega\right)+\frac{c}{\omega}F'_{C,l\beta}\left(\vec{k},\omega\right)e_{\alpha\gamma l}k_{\gamma} \tag{92}$$

From the other side, it would be interesting to find connections between the commonly used permittivity and permeability in both representations. From (56) we have:

$$R_{LL,\alpha\beta}\left(\vec{k},\omega\right)=\frac{\varepsilon_{LL,\alpha\beta}\left(\vec{k},\omega\right)-\delta_{\alpha\beta}}{4\pi} \tag{93}$$

From the other side, "C" representation possesses two forms, namely (75) and (76). Starting from the more widely used form (76), we get:

$$\begin{cases} R_{C,\alpha\beta}\left(\vec{k},\omega\right)=\dfrac{\varepsilon_{C,\alpha\beta}\left(\vec{k},\omega\right)-\delta_{\alpha\beta}}{4\pi} \\ \left[\mu_{\alpha\beta}\left(\vec{k},\omega\right)\right]^{-1}=\left[\delta_{\alpha\beta}+\dfrac{4\pi\omega}{c}F'_{C,\alpha\varsigma}\left(\vec{k},\omega\right)e_{\varsigma\gamma\beta}k_{\gamma}\left(\dfrac{1}{k_m k_m^*}\right)\right] \end{cases} \tag{94}$$

From the second equation of (94) one can express $F'_{C,\alpha\beta}\left(\vec{k},\omega\right)$ :

$$F'_{C,\alpha\beta}\left(\vec{k},\omega\right)=\frac{c e_{\beta l\gamma}k_l}{4\pi\omega}\left(\left[\mu_{\alpha\gamma}\left(\vec{k},\omega\right)\right]^{-1}-\delta_{\alpha\gamma}\right) \tag{95}$$

and finally get for the permittivity:

$$\varepsilon_{LL,\alpha\beta}\left(\vec{k},\omega\right)=\varepsilon_{C,\alpha\beta}\left(\vec{k},\omega\right)+\left(\frac{c}{\omega}\right)^2 e_{\alpha\gamma l}e_{\beta pm}k_{\gamma}k_p\left(\left[\mu_{lm}\left(\vec{k},\omega\right)\right]^{-1}-\delta_{lm}\right) \tag{96}$$

Following the same arguments, for the representation (75) we have:

$$\varepsilon_{LL,\alpha\beta}\left(\vec{k},\omega\right)=\varepsilon_{C,\alpha\beta}\left(\vec{k},\omega\right)+\frac{c}{\omega}e_{\alpha\gamma l}k_{\gamma}\left(\frac{c}{\omega}e_{lp\beta}k_{p}-\xi_{l\beta}\left(\vec{k},\omega\right)\right) \quad (97)$$

The permittivity in the “L&L” representation is proportional at least to the second order of the wave vector (quadratic spatial dispersion); $\xi_{l\beta}\left(\vec{k},\omega\right)\sim k$ in order to model magnetic response.

It has to be noted again, that in “L&L” representation there is only one function (actually, one family of functions) which fully determines the averaged optical response of media. From the other side, in “C” representation there are two functions (in both cases (75) and (76)). It is clear that from the known functions in “C” representation it is possible to construct one function in “L&L” representation, while the opposite transformation could not be done unambiguously.

The form (75) remains valid for any structures, while (76) can be used only in case when the permeability can be introduced – see (69), (70). Actually, the representation (75) is not only more general, but also is more convenient, because it fully reflects the physical nature of the processes, namely interaction of the metaatoms with the electric (not magnetic!) field.

It is believed, that the form (75) generally should be used in case when the basic processes causing magnetization are stipulated by the electric field; in other words, the form (75) has to be used in cases when there are no natural magnetic moments (like magnetic moments of natural atoms or molecules in ferromagnetic, for example), which can directly interact with magnetic fieldsat low frequencies.

After consideration of the phenomenological models of homogenization one can conclude that:

1. In both “L&L” and “C” representations it is possible to develop a phenomenological approach and reduce the homogenization procedure to several effective parameters with, in general, unknown functions/coefficients.
2. The effective parameters in general depend not only on the properties of media, but on the wave vector as well.
3. Due to the phenomenological nature of the presented here approach, it is in general impossible to separate in the effective parameters the parts which depend on the properties of media only from the parts, which contain dependence on the wave vector. Nevertheless, in case of a weak spatial dispersion (expansion of the respective functions up to the second order over the wave vector) it becomes possible to introduce effective material parameters in both cases of “L&L” and “C” representations.

### 2.1.4 Reduction to material equations for bianisotropic media in case of weak spatial dispersion

The presented here phenomenological approach results in a system of material equations in form (86), (87), namely:

$$\begin{cases} D_{C,\alpha}\left(\vec{k},\omega\right)=\left(\varepsilon^{(0)}_{C,\alpha\beta}\left(\omega\right)+\varepsilon^{(1)}_{C,\alpha\beta\gamma}\left(\omega\right)k_\gamma+\varepsilon^{(2)}_{C,\alpha\beta\gamma\delta}\left(\omega\right)k_\gamma k_\delta\right)E_\beta\left(\vec{k},\omega\right) \\ H_\alpha\left(\vec{k},\omega\right)=\left(\phi^{(0)}_{C,\alpha\beta}\left(\omega\right)+\phi^{(1)}_{C,\alpha\beta\gamma}\left(\omega\right)k_\gamma\right)E_\beta\left(\vec{k},\omega\right) \end{cases} \tag{98}$$

One can show that system (98) can be reduced to the form similar to the well-known and widely used in the literature material equations for bianisotropic media [Lindell 94]. In order to perform the necessary transformation, we use the known from the tensor algebra theorem which proves that any third-rank tensor can be presented as a sum of its symmetric and antisymmetric parts. The latter in turn can be presented as a tensor of the second rank multiplied by the Levi-Chivita tensor. Using this theorem, we can write:

$$\begin{cases} \varepsilon^{(1)}_{C,\alpha\beta\gamma}\left(\omega\right)=\varepsilon^{(1,sym)}_{C,\alpha\beta\gamma}\left(\omega\right)+\varepsilon^{(1,asym)}_{C,\alpha\beta\gamma}\left(\omega\right)=\varepsilon^{(1,sym)}_{C,\alpha\beta\gamma}\left(\omega\right)+G^{(\varepsilon)}_{C,\alpha p}\left(\omega\right)e_{p\beta\gamma} \\ \phi^{(1)}_{C,\alpha\beta\gamma}\left(\omega\right)=\phi^{(1,sym)}_{C,\alpha\beta\gamma}\left(\omega\right)+\phi^{(1,asym)}_{C,\alpha\beta\gamma}\left(\omega\right)=\phi^{(1,sym)}_{C,\alpha\beta\gamma}\left(\omega\right)+G^{(\phi)}_{C,\alpha p}\left(\omega\right)e_{p\beta\gamma} \end{cases} \tag{99}$$

Substituting (99) into (98) and taking into account that $e_{\alpha\beta l}k_\beta E_l=\frac{\omega}{c}B_\alpha$ , system (98) becomes:

$$\begin{cases} D_{C,\alpha}\left(\vec{k},\omega\right)=\left(\varepsilon^{(0)}_{C,\alpha\beta}\left(\omega\right)+\varepsilon^{(1,sym)}_{C,\alpha\beta\gamma}\left(\omega\right)k_\gamma+\varepsilon^{(2)}_{C,\alpha\beta\gamma\delta}\left(\omega\right)k_\gamma k_\delta\right)E_\beta\left(\vec{k},\omega\right)+G^{(\varepsilon)}_{C,\alpha\beta}\left(\omega\right)B_\beta\left(\vec{k},\omega\right) \\ H_\alpha\left(\vec{k},\omega\right)=G^{(\phi)}_{C,\alpha\beta}\left(\omega\right)B_\beta\left(\vec{k},\omega\right)+\left(\phi^{(0)}_{C,\alpha\beta}\left(\omega\right)+\phi^{(1,sym)}_{C,\alpha\beta\gamma}\left(\omega\right)k_\gamma\right)E_\beta\left(\vec{k},\omega\right) \end{cases} \tag{100}$$

The last form is rather close to the usually used Post [Lindell 94] form in case when the consideration is restricted to the first-order spatial dispersion $\varepsilon^{(2)}_{C,\alpha\beta\gamma\delta}\left(\omega\right)=0,\ \phi^{(1,sym)}_{C,\alpha\beta\gamma}\left(\omega\right)=0$:

$$\begin{cases} D_{C,\alpha}\left(\vec{k},\omega\right)=\left(\varepsilon^{(0)}_{C,\alpha\beta}\left(\omega\right)+\varepsilon^{(1,sym)}_{C,\alpha\beta\gamma}\left(\omega\right)k_\gamma\right)E_\beta\left(\vec{k},\omega\right)+G^{(\varepsilon)}_{C,\alpha\beta}\left(\omega\right)B_\beta\left(\vec{k},\omega\right) \\ H_\alpha\left(\vec{k},\omega\right)=G^{(\phi)}_{C,\alpha\beta}\left(\omega\right)B_\beta\left(\vec{k},\omega\right)+\phi^{(0)}_{C,\alpha\beta}\left(\omega\right)E_\beta\left(\vec{k},\omega\right) \end{cases} \tag{101}$$

It is important to emphasise that the final form (100) contains term $\varepsilon^{(1,sym)}_{C,\alpha\beta\gamma}(\omega)$ which manifests the fact that the first-order spatial dispersion has to be in general included in consideration as a separate term even in case of the Post material equations. In case of $\varepsilon^{(1,sym)}_{C,\alpha\beta\gamma}(\omega)=0$ system (101) takes the form basically equivalent to the Post equations for bianisotropic media:

$$\begin{cases} D_{C,\alpha}\left(\vec{k},\omega\right)=\varepsilon^{(0)}_{C,\alpha\beta}(\omega)E_\beta\left(\vec{k},\omega\right)+G^{(\varepsilon)}_{C,\alpha\beta}(\omega)B_\beta\left(\vec{k},\omega\right) \\ H_\alpha\left(\vec{k},\omega\right)=G^{(\phi)}_{C,\alpha\beta}(\omega)B_\beta\left(\vec{k},\omega\right)+\phi^{(0)}_{C,\alpha\beta}(\omega)E_\beta\left(\vec{k},\omega\right) \end{cases} \quad (102)$$

The parameters in (102) can be further investigated for reciprocal and non-reciprocal media, for example, applying the reciprocity theorem [Lindell 94], introducing the Tellegen parameter [Tellegen 48], etc.; these considerations are not done here. Note, that similar conclusion about necessity of introduction of additional terms in “traditional” material equations for bianisotropic media has been done in [Simovski 10] – see equations (28), (29) there. It is worth noticing again, that the extra term $\varepsilon^{(1,sym)}_{C,\alpha\beta\gamma}(\omega)$ in (101) appears in the first-order spatial dispersion, while the extra terms in [Simovski 10] correspond to the second order, and hence are responsible for different effects. It has to be also emphasised that the second-order spatial dispersion has to be in general taken into account in order to consider magnetic response (see [Petschulat 08]); in the frame of the phenomenological approach it corresponds to $\varepsilon^{(2)}_{C,\alpha\beta\gamma\delta}(\omega)\neq 0,\ \phi^{(1,sym)}_{C,\alpha\beta\gamma}(\omega)\neq 0$.

The first-order spatial dispersion term in (98), (99) could be compensated by the SFT – see (37). Nevertheless, it is easy to see, that SFT is able to compensate only the antisymmetric part $\varepsilon^{(1,asym)}_{C,\alpha\beta\gamma}(\omega)$, but not the symmetric part $\varepsilon^{(1,sym)}_{C,\alpha\beta\gamma}(\omega)$, which has to be in general retained in the material equations (101).

## 2.2. Multipole expansion (“C” representation)

### 2.2.1 Multipole approach

The multipole model was put forward in [Mazur 53], and later developed in a similar form in [Rusakoff 70]. The model is based on an averaging procedure using the Probability Distribution Function (PDF) for the positions and velocities of all charges, included in the consideration – statistical averaging, which is supposed to be equivalent to the originally assumed averaging over volume. Leaving alone the mathematical details of the model (which can be found in [Mazur 53]), here it is worth to recall the main ideas of the elaboration of the model.

The essence of the developed in [Mazur 53] averaging procedure is in summation of contributions from all atoms/molecules at the “Observation point” (see Fig. 1) using statistical math tools. In the framework of this approach each atom/molecule is considered as a cloud of positive and negative charges with some (a priori unknown) PDF over their coordinates and velocities. The condition $L_{\text{intra}} << L_{\text{inter}}$ allows us to use an expansion in the Taylor series of the potential, produced by each atom/molecule at the “Observation point”. As a result, the total contribution can be expressed in terms of averaged moments, namely the total charge of the system (zero-order moment), the electric dipole (first-order moment), the quadrupole and magnetic dipole (second-order moments), etc.

*It is worth noticing that the quadrupole and the magnetic dipole moments appear on the same level of Taylor expansion (second order) and therefore in general have the same order of magnitudes. In case of necessity to take into account a magnetic response of the atoms/molecules (as it takes place in case of MMs) it is usually necessary to use both moments together rather than voluntarily pick up just magnetic dipole moments and do not include into consideration the quardupole one. This requirement is stipulated by the fundamental principles of the multipole model and only in specific cases one can exclude the quadrupole moments of atoms/molecules from consideration.*

The model results in constructive expressions for $\vec{P}_C$ and $\vec{M}_C$ presented through the averaged dynamics of the charges in “C” form [Raab 05]:

$$\begin{cases} \vec{\mathrm{P}}(\vec{R},\omega) = \eta\left\langle \sum_{s}^{all\,charges} q_s \vec{r}_s \right\rangle - \nabla \bullet Q(\vec{R},\omega) \\ Q_{ij}(\vec{R},\omega) = \dfrac{\eta}{2}\left\langle \sum_{s}^{all\,charges} q_s r_{i,s} r_{j,s} \right\rangle \\ \vec{\mathrm{M}}(\vec{R},t) = \dfrac{\eta}{2c}\left\langle \sum_{s}^{all\,charges} q_s \left[ \vec{r}_s, \dfrac{\partial \vec{r}_s}{\partial t} \right] \right\rangle \end{cases} \tag{103}$$

The definitions clearly distinguish between microscopic (**r**) and macroscopic (**R**) coordinates, $q_k$ represents the charge, $N$ the total number of charges, and $\eta$ their density. The microscopic coordinates $\vec{r}$ designate the position vectors of the charges in a microscopic coordinate system, and $\vec{v}$ designate their velocities. The center of the microscopic coordinate system is chosen to be the center of symmetry of the charge distribution (consideration of the dependence on the origin of the coordinate system will be given later). The reason for the different coordinate systems derives from the averaging procedure for the averaged Maxwell equations [Ebbesen 98]. The microscopic coordinates are functions of the electric field and do not appear explicitly in the final expressions. Only one coordinate system, namely the *macroscopic system of coordinates* $\vec{R}$ (i.e., the space coordinate) remains.

The functions $\vec{D}(\vec{R},\omega)$ and $\vec{H}(\vec{R},\omega)$ (19), (20) contain electric dipole, electric quadrupole, and magnetic dipole contributions:

$$\begin{cases} \vec{D}(\vec{R},\omega) = \vec{E}(\vec{R},\omega) + 4\pi\vec{P}(\vec{R},\omega) \\ \vec{H}(\vec{R},\omega) = \vec{B}(\vec{R},\omega) - 4\pi\vec{M}(\vec{R},\omega) \end{cases} \tag{104}$$

$\vec{\mathrm{P}}(\vec{R},\omega)$, $Q_{ij}(\vec{R},\omega)$, and $\vec{M}(\vec{R},\omega)$ represent the electric polarization, the electric quadrupole tensor, and the magnetization, respectively. Capital letters $\vec{R}$ are used for macroscopic coordinates in the averaged Maxwell equations. The term

$$\nabla \bullet Q(\vec{R},\omega) = \frac{\partial Q_{ij}(\vec{R},\omega)}{\partial R_j} \tag{105}$$

is the divergence of the quadrupole tensor. It is necessary to take into account both the electric quadrupole and the magnetic dipole terms, because they are of the same order in the multipole expansion series [Jackson 75], [Raab 05].

It is important to realize that the formulas for the macroscopic polarization and magnetization are expressed in terms of the internal dynamics of the charges of the atoms/molecules, which are a priori functions of microscopic (not macroscopic!) fields. Even if we are able to write analytical forms for the dynamics, we will have to perform the averaging (see (103)) and express the functions $\vec{P}$ and $\vec{M}$ through the macroscopic fields. This finally closes the problem and makes from MEs a self-consistent system of equations, which can be (potentially) solved.

It should be emphasized that the multipole approach remains the only one, which allows us to create a logical connection from the microscopic to macroscopic forms of the MEs without any methodological gaps. The fact that finally this approach results in “C” form (one of the possible forms of MEs, obtained through the independent phenomenological consideration) serves as one more positive argument for the use of this model and its application to the problem of homogenization of MMs.

It has to be accepted that the basic conditions, under which system (103) has been elaborated are met for typical MMs in the optical domain rather poorly. Referring again to Fig. 1 and remembering typical experimental situations (for example, [Pshenay 09]), one can see that the distance between the metaatoms and the sizes of the metaatoms are of the same order, and the truncated Taylor expansion used in elaboration of (103) is not fully justified. Note that in contrast to the MM, in the case of natural materials system (103) works pretty well; the widely used dipole model for the permittivity is just the zero-order approximation of (103). Hence, the basic question about applicability of the multipole model to MMs remains open.

In spite of the fundamental doubts about its applicability, one can easily bring several arguments in favour of the multipole model:

1. The model offers a natural way to describe magnetization by introducing magnetic and quadrupole moments.
2. The model is physically clear and should be considered at least for the methodological reasons.
3. The model allows us to elaborate the functional forms for the introduced in phenomenological approach effective constants and fix the expressions for $\vec{P}$ and $\vec{M}$ as functions of the wave vector (in other words, find a functional form for spatial dispersion).
4. The model allows us to investigate the influence of the metaatom design on the optical properties of MMs.
5. The model allows us to investigate the influence of interactions between metaatoms on the optical properties of MMs.
6. The model allows us to investigate the influence of disorder (both spatial disorder in metaatom placements and disorder in eigen characteristics of the metaatoms) on the optical properties of MMs.
7. The model allows a natural extension beyond the purely plasmonic based metaatoms, for example to the case of combinations of plasmonic metaatoms and active quantum elements, or metaatoms consisting of purely quantum elements.

The multipole model created for MMs [Chipouline 11AP] contains parameters which can be tuned in order to compensate for the fundamentally stipulated discrepancies and finally fit the results of the model to the experimental and/or numerical data. It is believed, that the combination of the multipole approach with final tuning of these coefficients makes this model an extremely simple and versatile tool for investigation of optical properties of MMs [Petschulat 10]. In [Petschulat 08] analytical expressions for the effective permittivity and permeability have been elaborated for MMs based on double-wire structures. The charge dynamics has been treated using two coupled harmonic oscillator equations, possessing symmetric and antisymmetric oscillation modes, excited by the electric field. Using the expressions for the symmetric and antisymmetric modes, the dipole, quadrupole, and magnetic dipole terms (103) have been calculated as functions of the metaatoms and field parameters. Note, that in [Petschulat 08] only symmetric structures have been considered, extension to the case of asymmetric structures was performed in [Pshenay 11].

Below we will consider only 2D geometries (see Fig. 6), i.e., non-zero components of multipoles depend on the charge dynamics in the $(x, y)$ plane and can be written as:

$$\begin{cases} \left\langle \vec{j} \right\rangle_x (y,t) = -i\omega P_x + c\dfrac{\partial M_z}{\partial y} \\ P_x(y,t) = \eta \left\langle \sum_s^{all\,charges} e_s x_s \right\rangle - \dfrac{\partial Q_{xy}}{\partial y} \\ Q_{xy}(y,t) = \dfrac{\eta}{2} \left\langle \sum_s^{all\,charges} q_s x_s y_s \right\rangle \\ M_z(y,t) = \dfrac{\eta}{2c} \left\langle \sum_s^{all\,charges} q_s \left( x_s \dfrac{\partial y_s}{\partial t} - y_s \dfrac{\partial x_s}{\partial t} \right) \right\rangle \end{cases} \qquad (106)$$

Transition to the Fourier domain is not straightforward – both quadrupole and magnetic dipole moments depend nonlinearly on the coordinates. Nevertheless, it is assumed that to the first approximation any charge exhibits dynamics along just one direction: for example, in double wires it is the $x$ direction, in case of split rings in two arms the charges move along the $x$ direction and in the third arm – along the $y$ direction, etc. In case of simultaneous $(x, y)$ dynamics nonlinear response appears as a consequence of non-harmonic charge dynamics [Petschulat 09].

A particular geometry of double wires, shown in Fig. 6, allows a purely linear description; the electric field is polarized along the *x* axis, magnetic field is polarized along the *z* axis, and the wave propagates along the *y* axis. The wires are oriented along the *x* direction and are affected by the electric field which excites plasmonic modes in each wire; the coupled wires are considered as metaatoms, and multipole moments (103) are prescribed to these metaatoms.

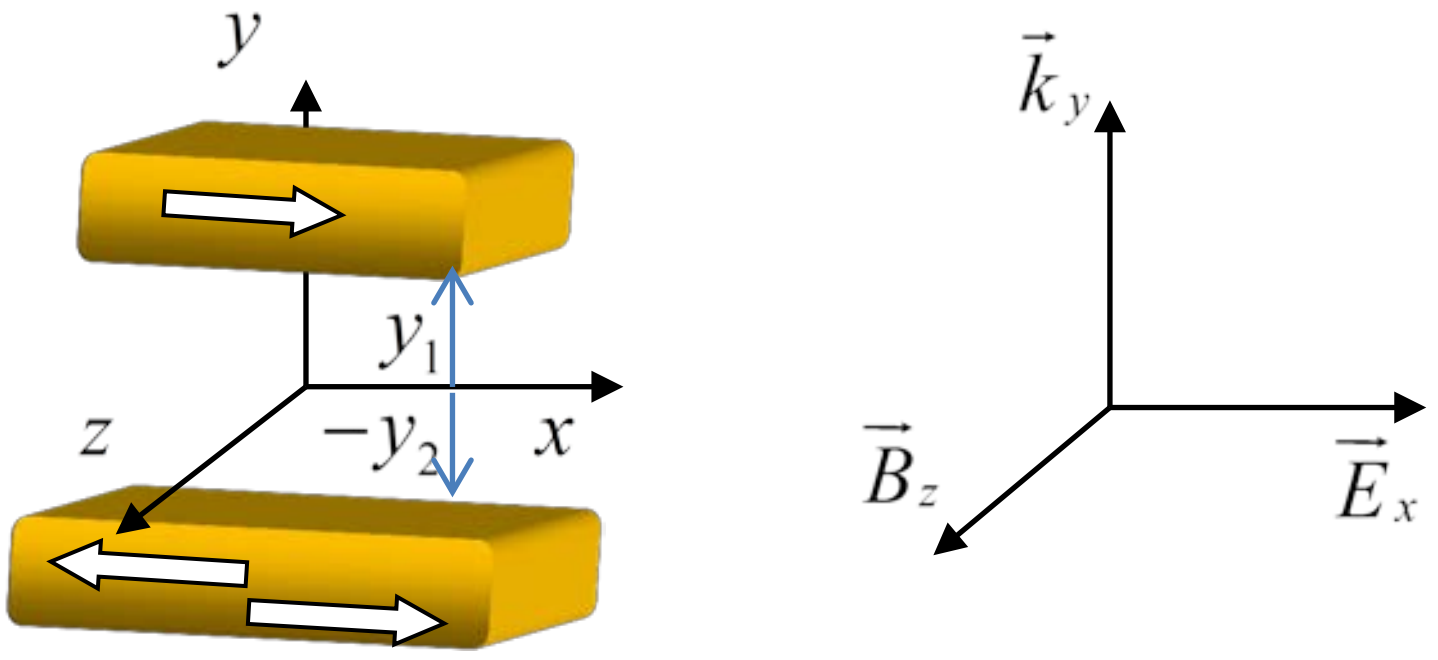

**Fig. 6: Double wire (in general case asymmetric) structure possessing dipole, quadrupole, and magnetic moments with respective direction of plane wave propagation and field polarizations. The system of microscopic coordinates is chosen so that its origin is placed in general at different distances $y_1$ and $y_2$ from both wires.**

In this case equations (106) can be further simplified (no dynamics along the *y* axis) and straightforwardly transformed into the frequency domain:

$$\begin{cases} \left\langle \vec{j} \right\rangle_x (y,\omega) = -i\omega P_x + c\dfrac{\partial M_z}{\partial y} \\ P_x(y,\omega) = \eta \left\langle \sum_s^{all\,charges} e_s x_s \right\rangle - \dfrac{\partial Q_{xy}}{\partial y} \\ Q_{xy}(y,\omega) = \dfrac{\eta}{2} \left\langle \sum_s^{all\,charges} q_s x_s y_{s,0} \right\rangle \\ M_z(y,\omega) = \dfrac{i\eta\omega}{2c} \left\langle \sum_s^{all\,charges} q_s x_s y_{s,0} \right\rangle \end{cases} \tag{107}$$

Note that $M_z(y,\omega) = \dfrac{i\omega}{c} Q_{xy}(y,\omega)$ and the magnetic dipole and the quadrupole moments have the same sign and are arranged together in the first equation in (107).

### 2.2.2 Dispersion relation elaboration

The derivation starts with the averaged (macroscopic) ME in "C" form in the frequency domain (19), (20). Here we consider MM consisting of double wire structures – one layer of such structure is presented in Fig. 2 on the top of a transparent substrate. The MM is formed by stacking of such layers in the *y* direction, and propagation of a plane wave along the *y* direction with the electric field polarized along the *x* direction is considered. This particular case illustrated in Fig. 2 is chosen in order to demonstrate main principles of the suggested model and avoid excessive math complications. The model is straightforwardly extendable to other metaatoms geometries, wave propagation directions, and field polarizations.

For a plane wave propagating along the positive *y* axis and with the electric field polarized along the *x* direction, the ME can be simplified and finally reduced to:

$$\begin{cases} P_x(y,\omega) = \eta \left\langle \sum_s^{all\,charges} e_s x_s \right\rangle - \dfrac{\partial Q_{xy}}{\partial y} \\ Q_{xy}(y,\omega) = \dfrac{\eta}{2} \left\langle \sum_s^{all\,charges} q_s x_s y_{s,0} \right\rangle \\ M_z(y,\omega) = \dfrac{i\eta\omega}{2c} \left\langle \sum_s^{all\,charges} q_s x_s y_{s,0} \right\rangle \end{cases} \qquad \begin{cases} \dfrac{\partial E_x}{\partial y} = -\dfrac{i\omega}{c} B_z \\ \dfrac{\partial H_z}{\partial y} = -\dfrac{i\omega}{c} D_x \\ D_x = E_x(y,\omega) + 4\pi P_{C,x}(y,\omega) \\ H_z = B_z - 4\pi M_{C,z} \end{cases} \tag{108}$$

which provides a self-consistent equation for the *x* component of the electric field $E_x(y,\omega)$ :

$$\frac{\partial^2 E_x(y,\omega)}{\partial y^2} + \frac{\omega^2}{c^2}\left(E_x(y,\omega) + 4\pi P_x(y,\omega)\right) + \frac{i4\pi\omega}{c}\frac{\partial M_z(y,\omega)}{\partial y} = 0 \tag{109}$$

It is interesting to note that the magnetic dipole and electric quadrupole contributions are identical. This is another proof that the electric quadrupole and magnetic dipole contributions have the same order of magnitude and both (not only the magnetic dipole term) have to be taken into account simultaneously [Mazur 53].

Equation (109) after Fourier transformation over the spatial coordinate $y$ becomes:

$$k_y^2 E_x(k_y,\omega) = \frac{\omega^2}{c^2}\left(E_x(k_y,\omega) + 4\pi P_x(k_y,\omega)\right) - \frac{4\pi k_y \omega}{c} M_z(k_y,\omega) \tag{110}$$

Assuming linear dependence of the multipole terms on the electric field (dependence on the magnetic field is negligible), one can write:

$$\begin{cases} P_x(k_y,\omega) = \left(p_x(k_y,\omega) - ik_y u_{xy}(k_y,\omega)\right) E_x(k_y,\omega) \\ Q_{xy}(k_y,\omega) = u_{xy}(k_y,\omega) E_x(k_y,\omega) \\ M_z(k_y,\omega) = m_x(k_y,\omega) E_x(k_y,\omega) \end{cases} \tag{111}$$

Substitution (111) into (110) results in the dispersion relation in the general form:

$$k_y^2 = \frac{\omega^2}{c^2}\left(1 + 4\pi p_x(k_y,\omega) - 4\pi i k_y u_{xy}(k_y,\omega)\right) - \frac{4\pi k_y \omega}{c} m_z(k_y,\omega) \tag{112}$$

It is also useful to write down the material equations, corresponding to the form (112):

$$\begin{cases} D_x(k_y,\omega) = \left(1 + 4\pi p_x(k_y,\omega) - i4\pi k_y u_{xy}(k_y,\omega)\right) E_x(k_y,\omega) \\ H_z(k_y,\omega) = \left(-\dfrac{k_y c}{\omega} - 4\pi m_z(k_y,\omega)\right) E_x(k_y,\omega) \end{cases} \tag{113.a}$$

$$\begin{cases} \varepsilon_x(k_y,\omega) = 1 + 4\pi p_x(k_y,\omega) - i4\pi k_y u_{xy}(k_y,\omega) \\ \mu_z(k_y,\omega) = \left(1 + \dfrac{4\pi\omega}{k_y c} m_z(k_y,\omega)\right)^{-1} \\ \xi_{zx}(k_y,\omega) = -\left(\dfrac{k_y c}{\omega} + 4\pi m_z(k_y,\omega)\right) \end{cases} \tag{113.b}$$

Dispersion relation (112) is the main result of this part and basically solves the problem of propagation of plane waves in media with higher multipoles. It is worth noticing that (112) is pretty universal – the charge dynamics and multipoles in (112) can be expressed based on classical, quantum, or semi-classical approaches and hence can be applied to extremely wide range of various problems. Moreover, this unified approach is highly desirable for education courses in the area of nanophotonics and electrodynamics of MMs, because it provides logical, internally consistent, allowing in most cases analytical treatment approach allowing deep understanding of physics of various problems using a single platform.

### 2.2.3 Physical interpretation of phenomenological coefficients

From the performed above consideration a clear and unambiguous connection between different types of excited modes in metaatoms (symmetric or antisymmetric) and the phenomenologically introduced in (86), (87) parameters can be revealed. It turned out [Petschulat 08], that the symmetric mode contributes to the dipole moment only, while the antisymmetric mode is responsible for appearance of the quadrupole and magnetic dipole terms, and, consequently, for the magnetic response of media. The antisymmetric mode can be excited in case of:

a) asymmetric structure of metaatoms (for example, in case of double wires it means different lengths of the wires - see [Pshenay 11] for details),
b) symmetric structure but inhomogeneous electric field (in [Petschulat 08] the retardation effect has been taken into account). In case of antisymmetric structure and symmetric external electric field the resulted quadrupole and magnetic moments do not contain wave vectors (are not spatially dispersive), while for the case of symmetric structure and inhomogeneous external electric field (for example, retarded field) the quadrupole and magnetic moments both depend on the wave vector, i.e. turn out to be spatially dispersive.

Comparison of the phenomenological (86), (87) and multipole (111) approaches allows us to prescribe clear physical interpretation of the coefficients in (86), (87) which requires more detailed metaatom structure consideration. For example, for symmetric double wires [Petschulat 08] and retarded at the scale of a metaatom field:

$$\begin{cases} p_x(k_y,\omega) = p_x(\omega) + p_x^{(2)}(\omega)k_y^2 \\ u_{xy}(k_y,\omega) = u_{xy}^{(1)}(\omega)k_y \\ m_z(k_y,\omega) = m_z^{(1)}(\omega)k_y \end{cases} \tag{114}$$

which gives for the coefficients of the phenomenological model (86), (87):

$$\begin{cases} \varepsilon^{(0)}_{C,\alpha\beta}(\omega) = 1 + 4\pi p_x(\omega) \\ \varepsilon^{(1)}_{C,\alpha\beta\gamma}(\omega) = 0 \\ \varepsilon^{(2)}_{C,\alpha\beta\gamma\delta}(\omega) = p^{(2)}_x(\omega) - i4\pi u^{(1)}_{xy}(\omega) \\ \phi^{(0)}_{C,\alpha\beta}(\omega) = 0 \\ \phi^{(1)}_{C,\alpha\beta\gamma}(\omega) = -\dfrac{c}{\omega} - 4\pi m^{(1)}_z(\omega) \end{cases} \qquad (115)$$

At the same time, for asymmetric structure and not retarded at the scale of the metaatom field:

$$\begin{cases} p_x(k_y,\omega) = p_x(\omega) \\ u_{xy}(k_y,\omega) = u^{(1)}_{xy}(\omega) \\ m_z(k_y,\omega) = m^{(1)}_z(\omega) \end{cases} \qquad (116)$$

$$\begin{cases} \varepsilon^{(0)}_{C,\alpha\beta}(\omega) = 1 + 4\pi p_x(\omega) \\ \varepsilon^{(1)}_{C,\alpha\beta\gamma}(\omega) = -i4\pi u^{(1)}_{xy}(\omega) \\ \varepsilon^{(2)}_{C,\alpha\beta\gamma\delta}(\omega) = 0 \\ \phi^{(0)}_{C,\alpha\beta}(\omega) = -4\pi m^{(1)}_z(\omega) \\ \phi^{(1)}_{C,\alpha\beta\gamma}(\omega) = -\dfrac{c}{\omega} \end{cases} \qquad (117)$$

Based on the given above consideration, it is rather straightforward to prescribe clear physical mean to the coefficients introduced in (86), (87), namely:

$\varepsilon^{(0)}_{C,\alpha\beta}(\omega)$ is the standard dielectric permittivity due to induced dipoles;

$\varepsilon^{(1)}_{C,\alpha\beta\gamma}(\omega)$ is the term corresponding to the appearance of the quadrupole moment (antisymmetric modes), where the quadrupole moment itself is not spatially dispersive. Physical situation is the metaatom with an antisymmetric mode, which could be excited by a homogeneous field (in this case the metaatom itself has to be asymmetric); retardation on the size of metaatom is not necessary in order to excite an antisymmetric mode of the metaatom. This term is responsible for bianisotropy [Simovski 10], [Kriegler 10] (in notations, accepted in these and similar papers).

$\varepsilon^{(2)}_{C,\alpha\beta\gamma\delta}(\omega)$ is the term corresponding to the appearance of the quadrupole moment (antisymmetric modes), where the quadrupole moment itself linearly depends on the wave vector (first-order spatial dispersion for the quadrupole moment) and dipole moment is proportional to the second order of the wave vector (second-order spatial dispersion for the dipole moment). Physically it corresponds to the situation when antisymmetric modes are excited by an inhomogeneous (for example, retarded) electric field; in this case the metaatom itself can be symmetric.

$\phi^{(0)}_{C,\alpha\beta}(\omega)$ is the term that basically is also responsible for the bianisotropy (in notations of [Simovski 10], [Kriegler 10]), and physically appears in case when the antisymmetric mode of the metaatoms can be excited by a homogeneous field (in this case the metaatom itself has to be asymmetric), and thus corresponds to $\varepsilon^{(1)}_{C,\alpha\beta\gamma}(\omega)$.

$\phi^{(1)}_{C,\alpha\beta\gamma}(\omega)$ is the term where extra contribution (the term $-\frac{k_y c}{\omega}E_x(k_y,\omega)$ is just the magnetic field) appearing in the case when the antisymmetric modes are excited by an inhomogeneous (for example, retarded) electric field.

It is worth noticing that the consideration of the problem here has been done based on the basic physical processes, appearing in the metaatoms, namely excitation of symmetric and antisymmetric modes. It is useful to summarize the coefficients with respect to these properties – see Table 2 below.

| **Symmetric properties of the metaatoms and fields** | **Type of excited modes** | **Respective coefficients** |
|---|---|---|
| Symmetric metaatoms, homogeneous electric field | Symmetric | $\varepsilon^{(0)}_{C,\alpha\beta}(\omega)$ |
| Asymmetric metaatoms, homogeneous electric field | Antisymmetric | $\varepsilon^{(1)}_{C,\alpha\beta\gamma}(\omega)$, $\phi^{(0)}_{C,\alpha\beta}(\omega)$ |
| Symmetric metaatoms, inhomogeneous electric field | Antisymmetric | $\varepsilon^{(2)}_{C,\alpha\beta\gamma\delta}(\omega)$, $\phi^{(1)}_{C,\alpha\beta\gamma}(\omega)+\frac{c}{\omega}$ |
| | | |

**Table 2.2: Symmetry properties of metaatoms and exciting fields, and the respective coefficients responsible for particular modes.**

It is interesting to note, that the developed approach establishes a way to determine the types of the modes exited in MM, provided the coefficients in Table 2.2 could be experimentally determined with the help of a retrieval procedure. In case of $\varepsilon^{(1)}_{C,\alpha\beta\gamma}(\omega)\sim 0,\ \phi^{(0)}_{C,\alpha\beta}(\omega)\sim 0$ one can conclude that the structure itself appears to be symmetric, and the magnetic response (in case $\phi^{(1)}_{C,\alpha\beta\gamma}(\omega)+\frac{c}{\omega}\neq 0$) is caused by a gradient of the electric field. In opposite case $\varepsilon^{(2)}_{C,\alpha\beta\gamma\delta}(\omega)\sim 0,\ \phi^{(1)}_{C,\alpha\beta\gamma}(\omega)+\frac{c}{\omega}\sim 0$ and respectively $\phi^{(0)}_{C,\alpha\beta}(\omega)\neq 0$ (non-zero magnetic response) the plasmonic oscillation mode is excited due to the asymmetry of the structure itself. The possibility to make a conclusion about microscopic processes (type of the excited oscillation mode) based on the macroscopic measurements (assuming that the mentioned above retrieval procedure can be designed) looks rather attractive and undoubtedly deserves further investigations.

### 2.2.4 Origin dependence of the multipole moments

It is known from the theory of multipoles (e.g., [Raab 05]) that the multipole moments, as they have been introduced in (103), depend on the origin of the microscopic system of coordinates (see the system of coordinates in Fig. 6). In fact, if the origin is shifted by a vector $\vec{g}(g_x,g_y,g_z)$ , the coordinates of all charges are shifted as well $\vec{r}_s=\vec{r}\,'_s+\vec{g}$ , and multipoles (103) become:

$$\begin{cases} \mathrm{P}_i(\vec{R},t)=\mathrm{P}'_i(\vec{R},t)-\dfrac{1}{2}\dfrac{\partial\left(d_i g_j+d_j g_i\right)(\vec{R},t)}{\partial R_j} \\ Q_{ij}(\vec{R},t)=Q'_{ij}(\vec{R},t)+\left(d_i g_j+d_j g_i\right)(\vec{R},t) \\ \mathrm{M}_i(\vec{R},t)=\mathrm{M}'_i(\vec{R},t)+e_{ijk}g_j\left\langle\vec{j}\right\rangle_k(\vec{R},t) \\ d_i(\vec{R},t)=\eta\left\langle\sum_s^{all\,charges}q_s\vec{r}_{s,i}\right\rangle \\ \left\langle\vec{j}\right\rangle_i(\vec{R},t)=\dfrac{\eta}{2c}\left\langle\sum_s^{all\,charges}q_s\vec{v}_{s,i}\right\rangle \end{cases} \tag{118}$$

Here $d_i$ are the components of the averaged dipole moment of the metaatom, $\left\langle \vec{j} \right\rangle_i$ are the components of the averaged current of the metaatom; it is also assumed that the metaatom is electrically neutral ( $\sum_s^{all\,charges} q_s = 0$ ). For the sake of simplicity only one particular metaatom geometry, presented in Fig. 6, is considered here. For this geometry (117) is simplified and becomes in the $\left(k_y, \omega\right)$ domain:

$$\begin{cases} \mathrm{P}_x(k_y,\omega) = \mathrm{P}'_x(k_y,\omega) - \dfrac{ik_y}{2} g_y d_x(k_y,\omega) \\ Q_{xy}(k_y,\omega) = Q'_{xy}(k_y,\omega) + \dfrac{1}{2} g_y d_x(k_y,\omega) \\ \mathrm{M}_z(k_y,\omega) = \mathrm{M}'_z(k_y,\omega) + \dfrac{i\omega}{2c} g_y d_x(k_y,\omega) \end{cases} \tag{119}$$

Substituting (118) into (109) we arrive to the conclusion that the dispersion relation depends on the origin in calculations of the multipoles:

$$\begin{aligned} k_y^2 E_x(k_y,\omega) &= \frac{\omega^2}{c^2}\Big(E_x(k_y,\omega) + 4\pi P_x(k_y,\omega)\Big) - \frac{4\pi k_y \omega}{c} M_z(k_y,\omega) = \\ &= \frac{\omega^2}{c^2}\Big(E_x(k_y,\omega) + 4\pi P'_x(k_y,\omega)\Big) - \frac{4\pi k_y \omega}{c} M'_z(k_y,\omega) - \frac{ik_y\omega^2}{c^2} g_y d_x \end{aligned} \tag{120}$$

The result of the origin dependence of the wave vector looks unacceptable, because of the wave vector obviously should not depend on the voluntary choice of the origin of the multipole calculations. It has to be emphasized, that anyway the origin dependence is a natural consequence of the elaborated in [Mazur 53] multipole model. Equations (118) and in our particular case of the double wires (119) are direct consequences of the recipe presented in [Mazur 53].

Let us take a look closer to this problem. The first row of the equation (120) is rigorous in terms of phenomenological approach: if we know exactly $P_x(k_y,\omega)$ and $M_z(k_y,\omega)$, then the wave vector can be found unambiguously (nevertheless, several solutions for the wave vector can exist). The question about origin dependence does not appear, because of both functions are assumed to be known without any approximations and irrespective to the multipole expansion. The question about origin dependence appears only when we start to apply the multipole expansion in order to elaborate an analytical form of the $P_x(k_y,\omega)$ and $M_z(k_y,\omega)$. The multipole expansion, nevertheless, is an approximation (especially if we restrict our consideration by quadrupole/magnetic dipole only) and gives the result with some limited accuracy: calculation of the multipole moments gives different results for the different origins, which is obvious. Hence, when we consider the problem of origin dependence, we always have to refer to the accuracy of the approximation we are working with.

The problem of the origin dependence roots to the problem of the field calculation from the known charge densities/dynamics at large enough distance – see [Landau&Lifshits 2]. The initial expression for the potential calculation is obviously origin invariant:

$$\varphi\left(\vec{R},t\right)=\sum_{k=1}^{N_{\max}}\frac{q_k}{\left|\vec{R}-\vec{r}_k\right|} \tag{121}$$

Here $q_k$, $\vec{r}_k$, $\vec{R}$ are the charge, the radius vector of the charge, and the radius vector of the observation point respectively, summation is going over the charge cloud up to the last charge with number $N_{\max}$.

Assuming that $\vec{R}>>\vec{r}_k$ we arrive to the standard multipole expansion, which is already not origin invariant:

$$\varphi\left(\vec{R},t\right)=\frac{1}{\left|\vec{R}\right|}\sum_{k=1}^{N_{\max}}q_k+\sum_{k=1}^{N_{\max}}q_k\vec{r}_{k,\alpha}\left(\frac{\partial}{\partial\vec{r}_\alpha}\frac{1}{\left|\vec{R}\right|}\right)+\sum_{k=1}^{N_{\max}}q_k\vec{r}_{k,\alpha}\vec{r}_{k,\beta}\left(\frac{\partial^2}{\partial\vec{r}_\alpha\vec{r}_\beta}\frac{1}{\left|\vec{R}\right|}\right) \tag{122}$$

Here, as usual, the first term is proportional to the total charge, and the second and third are the dipole and quadrupole approximation respectively. It is clear, that the origin invariance is lost due to the truncation of the expansion by the quadrupole approximation; in case of further expansion the origin invariance has to be recovered. The truncation at the quadrupole term is stipulated by a reasonability of this approximation: the quadrupole (and respectively magnetic dipole) term is the minimum required order of expansion which describes the magnetic effect due to current distribution. From the other side, this approximation is analytically treatable, and taking into account the next orders (octupoles etc.) makes the problem solvable only numerically; in this case introduction of the multipole approximation does not make too much sense. One can show that approximation (122) gives the result with the accuracy described by two scales. The first one is the relation between typical system size $a$ (size of the charge cloud) and the wavelength $\frac{\delta\varphi}{\varphi} \sim \left(\frac{a}{\lambda}\right)^2$ ; the second estimation is connected with the freedom of origin choice and can be estimated as $\frac{\delta\varphi}{\varphi} \sim \frac{a}{|R|}$; for the case of the metamaterial, considered here, $a$ is the typical size of the metaatom and $|R|$ is of the order of distance between the metaatoms. It has to be admitted, that the substitution of exact expression (121) by approximation (122) leads to the loss of the origin invariance, which on turn leads to the limited accuracy of (122). In the frame of the developed here model there is no rigorously justified methodology which could keep the origin invariance for the multipole (up to quadrupole/magnetic dipole order) approximation.

Moreover, it is easy to see, that the voluntarily choice of the origin leads to evidently non-physical results. Consider three systems of charges depicted in Figure 7. For the symmetric charge distribution the origin has to be chosen in the middle due to evident symmetry considerations. From the other side, if the origin is kept in the same point and the top dipole becomes shorter, then in the extreme case of negligible top dipole the quadrupole moment has to be zero, which takes place only if the origin appears between the charges (dashed *x* axe in Fig. 7 right picture).

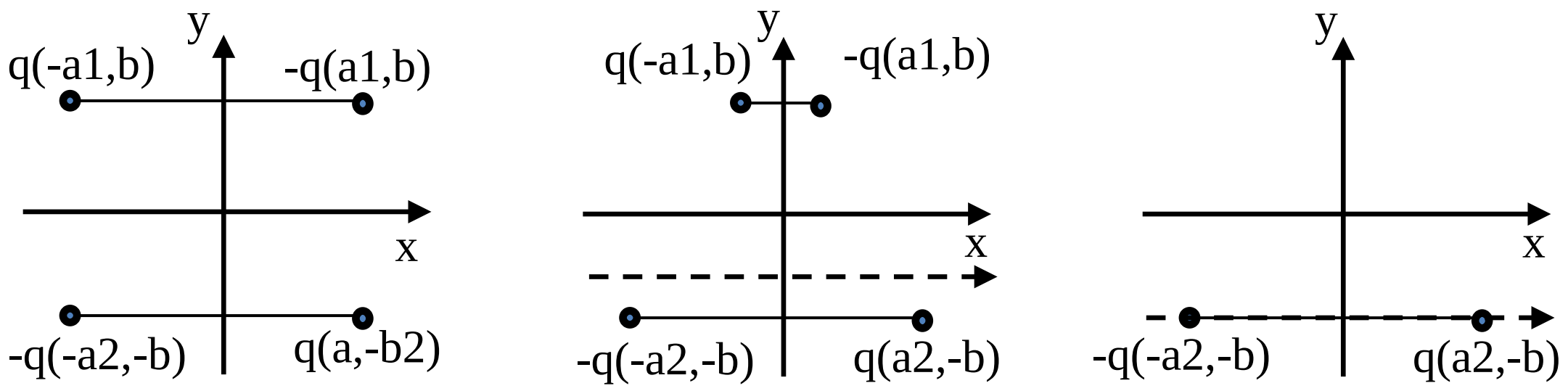

**Fig. 7: Schematic representation of three charge distribution with the same position of the origin (solid axes) and variable *x* axe (dashed *x* axe). For the symmetric charge distribution (left picture) the origin has to appear in the middle due to the symmetry consideration; the same origin position (solid *x* axe) for a single dipole (right picture) leads to unphysical appearance of magnetic response, while the origin for the dipole has to lay between the charges (dashed *x* axe). This consideration gives a hint to the choice for the origin position for the intermediate charge system (middle picture).**

It becomes intuitively clear that in the middle case the origin has to be chosen lower towards to the bigger dipole (dashed *x* axe), which allows us to suggest new approach to the problem of origin for the multipole expansion – instead of require rather questionable condition of the origin independence (which does not appear as a natural consequence of the developed model) we apply an extra condition, which fixes the origin for any system based on its charge distribution.

The origin can be fixed by rather physically evident requirement of zero quadrupole and magnetic dipole moments in the case when only the symmetric oscillation mode is excited (or, better to say, the antisymmetric mode is not excited). In fact, in case of symmetric oscillations there are no circular currents und hence there is no reason for the appearance of the magnetic response; from the other side, non-zero quadrupole and magnetic dipole moments would mean that this magnetic response takes place. It is enough to consider the quadrupole moment only (the magnetic moment can be considered in the same way). Let us assume (see Fig. 6) that the quadrupole moment is calculated in some coordinate system:

$$Q_{xy}(k_y,\omega)=\frac{1}{2}\eta\left[q_1 y_1 x_1\left(k_y,\omega\right)-q_2 y_2 x_2\left(k_y,\omega\right)\right] \tag{123}$$

here $q_{1,2}, y_{1,2}, x_{1,2}$ are the charges, the positions over $y$, and the deviation from an equilibrium position of the charges in the upper and lower wires, respectively. In another system of coordinates the new quadrupole moment is given by:

$$Q'_{xy}(k_y,\omega)=Q_{xy}(k_y,\omega)+\frac{1}{2}g_y d_x=\frac{1}{2}\eta\left[q_1 y_1 x_1\left(k_y,\omega\right)-q_2 y_2 x_2\left(k_y,\omega\right)\right]+\frac{1}{2}g_y\eta\left[q_1 x_1\left(k_y,\omega\right)+q_2 x_2\left(k_y,\omega\right)\right]$$
(124)

In case of the symmetric mode $x_1\left(k_y,\omega_{sym}\right)=x_2\left(k_y,\omega_{sym}\right)$ and expression for the quadrupole moment becomes:

$$Q'_{xy}(k_y,\omega)=\frac{1}{2}\eta\left(\left[q_1 y_1-q_2 y_2\right]+g_y\left[q_1+q_2\right]\right)x_2\left(k_y,\omega_{sym}\right)=0 \tag{125}$$

which leads to the fixed position for the origin of the coordinates, at which the quadrupole moment (and, consequently, the magnetic response) is zero:

$$g_y = \frac{q_2 y_2 - q_1 y_1}{q_1 + q_2} \tag{126}$$

Hence, at least for the considered here particular case, the origin of the system of coordinates can be unambiguously fixed. Note that for the case of identical wires $q_1 = q_2$ and the origin has to be placed in the middle point $y_1 = y_2$; for the absence of one of the wires (right picture in fig. X) transformation (126) shifts the origin to another one and places the origin such way that the quadrupole/magnetic dipole moments disappear.

In the presented here model the averaged current (107) and final expression for the wave vector (120) contain the same combination of $P_x(k_y,\omega)$ and $M_z(k_y,\omega)$, namely $\frac{\omega}{c}P_x(k_y,\omega) - k_y M_z(k_y,\omega)$. It means, that the origin independence of the wave vector is equivalent to the origin independence for the averaged current, which cannot be logically justified in case of multipole expansion; the only requirement which could be utilized is that the accuracy of the accepted approximation should not degrade at the choice of different origins.

The extra terms for the quadrupole and magnetic dipoles, appearing due to the origin shift (see (119)) cannot cancel each other due to the relation between them $Q_x(k_y,\omega) = \frac{i\omega}{c} M_z(k_y,\omega)$ and the fact that both $Q_x(k_y,\omega)$ and $M_z(k_y,\omega)$ appear in both averaged current and wave vector expression, as it was already mentioned, in the same combination $-\frac{ik_y\omega}{c}Q_x(k_y,\omega) - k_y M_z(k_y,\omega) = -2\frac{ik_y\omega}{c}Q_x(k_y,\omega)$.

One more question to be considered here is the relation of the origin dependence and the SFT. According to the general rules, the SFT (37) for the considered here double wires can be written as (compare with (119)):

$$\begin{cases} \mathrm{P}_x(k_y,\omega) = \mathrm{P}'_x(k_y,\omega) - \frac{i\omega}{4\pi c}T_{1,x} + ik_y T_{2,z} \\ \mathrm{M}_z(k_y,\omega) = \mathrm{M}'_z(k_y,\omega) + \frac{ik_y}{4\pi}T_{1,z} + \frac{i\omega}{c}T_{2,z} \end{cases} \tag{127}$$

In order to compensate for the shift, both transformations have to be equivalent, namely:

$$\begin{cases} -\dfrac{ik_y}{2}g_y d_x(k_y,\omega) = -\dfrac{i\omega}{4\pi c}T_{1,x} + ik_y T_{2,z} \\ +\dfrac{i\omega}{2c}g_y d_x(k_y,\omega) = \dfrac{ik_y}{4\pi}T_{1,z} + \dfrac{i\omega}{c}T_{2,z} \end{cases} \tag{128}$$

The last requirement (128) can be satisfied by unlimited number of variants for $T_{1,x}$ and $T_{1,z}$. From the other side, (128) cannot be satisfied if both $T_{1,x}$ and $T_{1,z}$ are zeros, in this case:

$$\begin{cases} \mathrm{P}_x(k_y,\omega) = \mathrm{P}'_x(k_y,\omega) + ik_y T_{2,z} \\ \mathrm{M}_z(k_y,\omega) = \mathrm{M}'_z(k_y,\omega) + \dfrac{i\omega}{c}T_{2,z} \end{cases} \tag{129}$$

$$\begin{cases} -\dfrac{ik_y}{2}g_y d_x(k_y,\omega) = ik_y T_{2,z} \\ +\dfrac{i\omega}{2c}g_y d_x(k_y,\omega) = \dfrac{i\omega}{c}T_{2,z} \end{cases} \tag{130}$$

and there is no nontrivial solution for $T_{2,z}$. Transformations (130) do not change dispersion relation (110), and cannot compensate for the origin shift (119). Using transformation (130) it is possible either compensate for the origin shift in quadrupole or in magnetic dipole contribution, but not for both of them simultaneously.

An ability to compensate for the origin shift, given by nonzero solutions $T_{1,x}$ and $T_{1,z}$ of (128) means that final field solution has to be transformed as well according to (37). Basically, it does not bring any news for the origin dependence problem because of it does not fix the origin based on some extra conditions. As it was shown above (see 1.5 "Serdyukov-Fedorov transformation"), the nonzero $\vec{T}_1$ means solution for some other charge/current distributions; in the case of origin shift the charge/current are formally changed (due to the limited accuracy of the multipole approximation), and the Serdyukov-Fedorov transformation compensates for this change.

One can finally conclude, that:

1. In the frame of the developed here approach the origin dependence appears as a consequence of the limited accuracy of the multipole approximation, which is estimated to be $\sim\dfrac{a}{|R|}$, here $a$ is the typical size of the metaatom and $|R|$ is of the order of distance between the metaatoms.
2. The origin independence cannot be designed in the frame of the developed here model; moreover, the requirement of this independency seems to be rather questionable.
3. In order to fix the problem, we have introduced another requirement, that fixes the origin for each metaatom and which clearly leads to reasonable limiting cases.

4. Consideration of this problem in the frame of the presented approach requires further investigation.

The conclusions of this part evidently contradict to the accepted in the literature ones that the multipole model can be constructed such way, that the requirement of the origin independency can be satisfied [Raab 05]. Note that in the developed in [Raab 05] approach initial expressions for the multipoles differ from ones, elaborated in [Mazur 53] and which our consideration is based on. The main difference between our consideration and the theory in [Raab 05] is in following. In [Raab 05] the multipole expansion (i.e. expansion of the charge dynamics) is mixed with the expansion of the local field; in our model the field expansion and consequent consideration of the local/averaged fields and the consideration of their mutual relations is not necessary for the presented above conclusions. At the moment the relation of the developed here model and one in [Raab 05] requires further investigation.

## 2.3 Introducing of Effective Parameters

### 2.3.1 Elaboration of Effective Parameters

The introduced above scheme allows us to determine unambiguously the terms "effective parameter" and some other notations, which appeared in the literature in the contents of the homogenization procedure. Most generally, the effective parameters are functions which connect introduced at the first step of the homogenization procedure $\langle\rho\rangle$ and $\langle\vec{j}\rangle$ with the averaged electric and magnetic fields $\vec{E}$ and $\vec{B}$. Usually the relations between $\langle\rho\rangle$ and $\langle\vec{j}\rangle$ and $\vec{E}$ and $\vec{B}$ are not used; instead the relations between $\vec{P}$ and $\vec{M}$ (in any representations) and $\vec{E}$ and $\vec{B}$ are under consideration. It has been shown above, that in the case of the "L&L" representation it is possible to introduce the effective parameter (effective permittivity) $\varepsilon_{LL,\alpha\beta}\left(\vec{k},\omega\right)=\delta_{\alpha\beta}+4\pi R_{LL,\alpha\beta}\left(\vec{k},\omega\right)$ (56), which connects $\vec{D}$ and $\vec{E}$. The effective permittivity is a function of material properties and is spatially dispersive (depends on wave vector); in other words, is nonlocal.

The nonlocality (spatial dispersion) caused a discussion in literature whether it makes sense to call $\varepsilon_{LL,\alpha\beta}\left(\vec{k},\omega\right)$ an effective *material* parameter. Following [Menzel 08], the effective permittivity $\varepsilon_{LL,\alpha\beta}\left(\vec{k},\omega\right)$ is an effective parameter, but is not an effective *material* parameter because it includes information not only about material, but also depends on the wave vector, in other words contains information about external electromagnetic fields. Similar considerations have been published in [Simovski 11] where also two different notations – Characteristic Material Parameters (CMP) and Effective Material Parameters (EMP) – have been suggested.

Here the question about necessity of introducing of new paradigm for the effective parameters is considered in view of the presented in Section 2.2 consideration based on the types of modes excited in metaatoms.

In order to create a consistent terminological basis, it has to be mentioned, that the introduced effective parameters in case of strong spatial dispersion (for example, the permittivity in case of the "L&L" representation) solves the problem of homogenization, and to this extend does not require any other comments or discussions. If the functional form of $\varepsilon_{LL,\alpha\beta}\left(\vec{k},\omega\right)$ is found, then this function contains information about material properties (for example, eigen modes of the electron oscillations in the metal nanoresonators), and the excitation conditions (wave-vector dependence). In general, both properties – eigen modes and excitation conditions – are not separable.

The physical meaning of the effective constants, introduced in the homogenization procedure, should be properly appreciated. The effective constants appear in the model (phenomenological or multipole) as the functions describing charge dynamics. It is necessary to recall that the microscopic MEs consist of not only field equations, but also include equations for the charge dynamics – see the last equation in (1). This equation (and all information about charge dynamics) is lost in the homogenization procedure; in other words, this equation is not averaged in the homogenization. It is clear, that the information contained in this equation is also lost and has to be somehow compensated, which is the basic reason for the multiple forms of various homogenization models (it has to be noted, that this equation has been basically kept in [Petschulat 08], which is the reason for pretty straightforward elaboration of the model). Instead of the systematic consideration of the charge dynamics, the phenomenological approach just gives some frames for the homogenization model, resulting in some expressions connecting averaged charge density and current and averaged fields. The coefficients between them (effective parameters) should contain information about charge dynamics caused by the averaged fields.

At this stage it has to be clearly realized, that these coefficients contain information about charge dynamics in a particular excitation situation, and not only about the material properties. The difference becomes clear if we again recall the symmetric and antisymmetric modes in, for example, double wires. The existence of the modes itself is the characteristic of the system and does not depend on the external fields. But the type of the excited mode can depend (but not necessarily!) on the external field structure – see Table 1 where it is summarized that the antisymmetric mode can be excited due to both asymmetry of the structure and inhomogeneity of the field (asymmetry of the field distribution). In general, the modes which are excited and which determine the material response are functions of both material properties and properties of the external fields. It is possible to create metaatoms with modes determined by the properties of the metaatoms only – asymmetric structures (for example, double wires with different lengths closely placed to each other – see [Menzel 12]) with the sizes much smaller than the wavelength. In this case the coefficients $\varepsilon_{C,\alpha\beta}^{(0)}(\omega)$, $\varepsilon_{C,\alpha\beta\gamma}^{(1)}(\omega)$ and $\phi_{C,\alpha\beta}^{(0)}(\omega)$ form material equations, which nevertheless describe spatial dispersion of the first order. This clearly shows that the presence/absence of spatial dispersion in material response is not an indicator of the physical properties of the media, but rather a manifestation of interactions of charges with the fields and the charge dynamics. If all five coefficients in Table 1 are not zero (the antisymmetric mode is excited due to both the structure and field asymmetries), the situation is basically the same, but the material equations contain spatial dispersion of the second order as well.

The presented above consideration shows that the spatial dispersion itself can hardly be considered as a criterion for introduction of some new notations. It would be more physically justified to consider the effective parameters based on the types of the modes in particular structures under particular excitation conditions. For example, following the results summarized in Table 1, one can subdivide the magnetic response in metaatoms (and the respective MM) into two categories: the first one where the antisymmetric modes appear as a result of the asymmetry of the structure itself, and the second one, where the antisymmetric modes appear as a result of the asymmetric excitation. For both types the coefficients in Table 1 can be considered as material effective parameters, because they all depend on the material properties and do not depend on the wave vector.

The MM of the first and second types can be rather easily distinguished experimentally.

If the retrieval procedure based on (99) will be applied, it will be clear which type of structure has been tested: for the first type the coefficients $\varepsilon_{C,\alpha\beta\gamma\delta}^{(2)}(\omega)$, $\phi_{C,\alpha\beta\gamma}^{(1)}(\omega)$ are expected to be close to zero, and for the second type the coefficients $\varepsilon_{C,\alpha\beta\gamma}^{(1)}(\omega)$, $\phi_{C,\alpha\beta}^{(0)}(\omega)$ should be about zero. The information about relative contribution of the coefficients would reveal information about basic physical processes in the metaatoms and undoubtedly enhance our insight about electrodynamics of the MM. Moreover, this method allows us to extract *microscopic* information based on *macroscopic* measurements of the effective response of the MM, which makes this approach an extremely useful tool in the lab.

### 2.3.2 Impossibility of unambiguous Effective Parameters determination for bulk materials

It has to be realized that the problem of determination of Effective Parameters for bulk materials does not make much practical sense. From the theoretical point of view, the only problem which can be stated and (potentially) solved for the bulk materials is finding the dispersion relation, which does not mean even introduction of the effective parameters. The effective parameters can be introduced as some coefficients between polarizability and magnetization and the electric and magnetic fields, but it can NOT be done unambiguously – see again SFT (37), which give birth of unlimited number of different effective parameters, each set of them nevertheless satisfy MEs.

In fact, most authors do not look for the step by step elaboration of the model, describing the averaged characteristics, but just postulate the relations between polarizability, magnetization and the electric and magnetic fields and then find, for example, dispersion relation - see Fig. 7, which illustrates different ways of possible elaboration of the homogenization models. It is clear, that the chain "Microscopic Maxwell equations" to "Bulk material – introduction of ε, μ", shown by the red vertical arrow, is not a natural way of the elaboration of the homogenization model, but rather an attempt to avoid the detailed and consequent consideration.

A functional form of the material equations has to be elaborated based on the phenomenological and /or multipole approaches, which has been done for the bulk materials in this work. We argue, that the presented here consideration covers all possible forms of the material equations in the frame of the commonly used paradigm of spatial dispersion.

### 2.3.3 Effective Retrieved Parameters and their relation to the Effective Parameters

The problem of the effective parameters retrieval is to some extend out of the scope of the presented consideration. The reason is that in the optical domain due to the high level of losses the vast majority of tests have been performed not with bulk MMs (which is the main object of the presented consideration) but rather with metasurfaces [Vinogradov 11], [Moritz 10], where only several layers of metaatoms are stacked together. In this case the presented above consideration is not directly applicable. For example, expressions (53), (54), and (61) could not form the basis for further consideration (nevertheless, all the conclusions made in Part 1 about different representations and their mutual transformations remain valid). This problem of fundamental inapplicability of the "bulk expressions" to metasurfaces are frequently mentioned, but very rearly properly analyzed; anyway, all the presently accepted approaches for the effective parameter retrieval are reduced to the "bulk expressions" in one or another form – see, for example, [Albooyeh 11]. The theoretical background for the retrieval procedures is based on some hypothesis about relations between polarizability and magnetization on the averaged field. In its original form the retrieval procedure assumed very simple forms of the relations for polarizability and magnetization [Weir 74], [Nicholson 70], [Smith 02], [Smith 05] which did not include bianisotropy and corresponded in terms of (89) to

$$\begin{pmatrix} \varepsilon^{(0)}_{C,\alpha\beta}(\omega) & 0 & 0 \\ 0 & \phi^{(1)}_{C,\alpha\beta\gamma\delta}(\omega) & 0 \end{pmatrix} \tag{131}$$

Bianisotropy has been included into consideration in [Silveirinha 07] and in the retrieval algorithm of [Chen 05], which corresponds in terms of the developed here approach to the form:

$$\begin{pmatrix} \varepsilon^{(0)}_{C,\alpha\beta}(\omega) & \varepsilon^{(1)}_{C,\alpha\beta\gamma\delta}(\omega) & 0 \\ \phi^{(0)}_{C,\alpha\beta}(\omega) & \phi^{(1)}_{C,\alpha\beta\gamma\delta}(\omega) & 0 \end{pmatrix} \tag{132}$$

It is worth noticing again, that the developed in this publication approach gives a different representation for the retrieval procedure (98) which includes not only bianisotropy, but also excludes the influence of the magnetic field on the effective parameters.

The retrieved from the reflection and transmission measurements parameters (whatever representation it is based on) are the so called “effective refractive index” and the “effective surface impedance”. The main goal of the theoretical models is to establish a reasonable correspondence between the retrieved from the measured data effective refractive index and impedance and parameters of the developed model. Often it is assumed that the effective parameters of the metasurfaces can be described by the same type of relations as for bulk MMs - for example, coefficients in (98) or (100), or the respective coefficients in some other representation.

## 2.5. Conclusion of Part 2

The developed in this chapter scheme is summarized in Fig. 8. Developing of the averaging procedure results in roughly three levels of MEs: the first level is the microscopic MEs (the starting level), the second level are the different representations for macroscopic MEs, where polarizability and magnetization $\vec{P}$ and $\vec{M}$ can be introduced, and the third level where the homogenization problem results in the dispersion relation for bulk materials and in a surface impedance and an effective refractive index for layered materials. Possible transitions between the first and second levels are represented by the phenomenological route and the multipole model; it is worth reminding that the multipole model arrives to the same “C” form of MEs representation.

The effective parameters in general turn out to be functions of the wave vector and hence cannot be called the effective material parameters. In order to separate the effect of spatial dispersion, the effective parameters can be expanded into series over the wave vector, and the appearing coefficients can be accepted as effective material parameters.

A new form of material equations has been suggested in the frame of the phenomenological approach in the case of weak (up to the second order) dispersion. Comparison between the phenomenological and multipole approaches based on the symmetry consideration allowed us to prescribe a clear physical meaning to all introduced material parameters. The use of the retrieval algorithm based on the suggested form of the material equations would allow us to determine the types of the internal charge mode dynamics, on which the response of MMs is based.

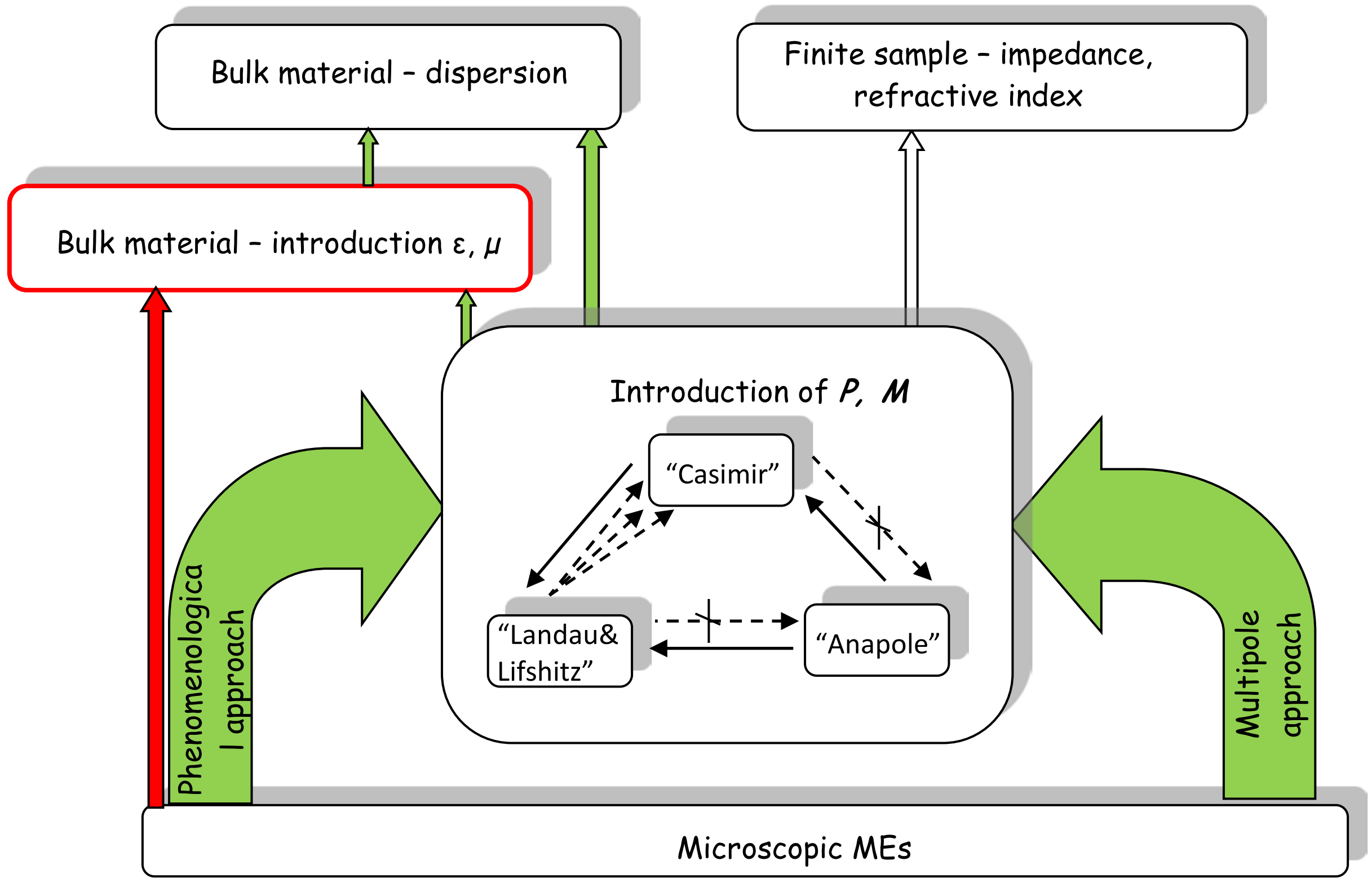

**Fig. 8: Schematic representation of different levels of the averaging procedure. "Red way" - it's NOT a homogenization, but rather a way to avoid elaboration of the homogenization model.**